\documentclass[10pt,twocolumn,twoside]{IEEEtran}
\usepackage{graphicx}
\usepackage{amsmath}
\usepackage{amssymb}
\usepackage{array}
\UseRawInputEncoding
\newcommand{\PreserveBackslash}[1]{\let\temp=\\#1\let\\=\temp}
\newcolumntype{C}[1]{>{\PreserveBackslash\centering}p{#1}}
\newcolumntype{R}[1]{>{\PreserveBackslash\raggedleft}p{#1}}
\newcolumntype{L}[1]{>{\PreserveBackslash\raggedright}p{#1}}
\usepackage[usenames]{color}
\usepackage{colortbl,booktabs}
\usepackage{tabularx}
\usepackage{booktabs}
\usepackage[subfigure]{tocloft}
\usepackage{subfigure}
\usepackage{bm}
\usepackage{cite}
\usepackage{balance}

\usepackage{algorithm}
\usepackage{hyperref}
\usepackage{threeparttable}
\usepackage{amsmath}
\usepackage{dcolumn}
\usepackage{multirow}
\usepackage{booktabs}
\usepackage{amsfonts}
\usepackage{framed}
\allowdisplaybreaks[3]
\hypersetup{hidelinks, 
	colorlinks=true,
	allcolors=black,
	pdfstartview=Fit,
	breaklinks=true}

\newcolumntype{d}[1]{D{.}{.}{#1}}
\begin{document}

\bibliographystyle{IEEEtran} 
\title{Near-Field Channel Estimation \\in Mixed LoS/NLoS Environments \\ for Extremely Large-Scale MIMO Systems}
\author{{Yu Lu,~\IEEEmembership{Student Member,~IEEE}, and Linglong Dai,~\IEEEmembership{Fellow,~IEEE}}
		
\thanks{All authors are with the Department of Electronic Engineering, Tsinghua University, Beijing 100084, China, and also with the Beijing National Research Center for Information Science and Technology (BNRist), Beijing 100084, China. (e-mails: y-lu19@mails.tsinghua.edu.cn, daill@tsinghua.edu.cn).}

\thanks{This work was supported in part by the National Key Research and Development Program of China (Grant No. 2020YFB1807201) and in part by the National Natural Science Foundation of China (Grant No. 62031019), and in part by the European Commission through the H2020-MSCA-ITN META WIRELESS Research Project under Grant 956256. (\textit{Corresponding author: Linglong Dai.})}
}
\maketitle
\vspace{-0mm}
\begin{abstract}
 
Accurate channel model and channel estimation are essential to empower extremely large-scale MIMO (XL-MIMO) in 6G networks with ultra-high spectral efficiency. With the sharp increase in the antenna array aperture of the XL-MIMO scenario, the electromagnetic propagation field will change from far-field to near-field. Unfortunately, due to the near-field effect, most of the existing XL-MIMO channel models fail to describe mixed line-of-sight (LoS) and non-line-of-sight (NLoS) path components simultaneously. In this paper, a mixed LoS/NLoS near-field XL-MIMO channel model is proposed to match the practical near-field XL-MIMO scenario, where the LoS path component is modeled by the geometric free space propagation assumption while NLoS path components are modeled by the near-field array response vectors. Then, to define the range of near-field for XL-MIMO, the MIMO Rayleigh distance (MIMO-RD) and MIMO advanced RD (MIMO-ARD) is derived. Next, a two stage channel estimation algorithm is proposed, where the LoS path component and NLoS path components are estimated separately. Moreover, the Cram\'er-Rao lower bound (CRLB) of the proposed algorithm is derived in this paper. Numerical simulation results demonstrate that, the proposed two stage scheme is able to outperform the existing methods in both the theoretical channel model and the QuaDRiGa channel emulation platform.

\end{abstract}

\vspace{-1mm}

\section{Introduction}\label{S1}
Due to the emergence of new applications such as digital twin, holographic imaging, and extended reality, the spectral efficiency of 6G is expected to grow tenfold~\cite{Ray_6G,RAY6G,Near_field_Mag}. {\color{black}To achieve higher spectral efficiency, the extremely large-scale multiple-input multiple-output (XL-MIMO) is regarded as one of the most important technologies for 6G~\cite{XLMIMO,XL_MIMO2}.} Compared with the massive MIMO technology for 5G, the sharp increase of antenna aperture in XL-MIMO for 6G induces the fundamental change of the electromagnetic field property. The electromagnetic field can be divided into the far-field region and the near-field region. {\color{black}The boundary between these two regions is the Rayleigh distance (RD) $ Z = 2D^2/\lambda $~\cite{Ray}, which is defined in a multiple-input single-output (MISO) scenario.} 
The RD is proportional to the square of the array aperture $ D $ and the inverse of wavelength $ \frac{1}{\lambda } $. {\color{black}Since the antenna number in massive MIMO systems for 5G is not large enough (e.g., 64 antennas
in~\cite{massiveMIMO}), the near-field range of up to a few tens of meters is negligible. For example, with the carrier frequency at 28 GHz, the RD for an array with array aperture $ D $ as 0.1 $ \rm m $ is only 1.9 $\rm m $~\cite{RD_example}.} Thus, users are likely located in the far-field region, where the far-field channel can be modeled with planar wavefronts assumption. {\color{black}On the contrary, as the antenna number dramatically increases in XL-MIMO systems (e.g., 1024 antennas in~\cite{numberXLMIMO}), the near-field range will expand by orders of magnitude, which can be up to several hundreds of meters.} For instance, with the array aperture as 1 $ \rm m $ at 28 GHz, the RD is 187 $ \rm m $, which nearly covers a region of a cellular cell. In this case, the XL-MIMO channel should be modeled under near-field assumption with spherical wavefronts.

{\color{black}In XL-MIMO communications, the transmitter equipped with an extremely large-scale antenna array (ELAA) serves multiple users by exploiting the high spatial multiplexing gain.} Except for antenna array, reconfigurable intelligent surface (RIS) is one of the main implementation methods of XL-MIMO~\cite{Near_field_Mag}, which is an energy efficient alternative of ELAA.
In order to obtain spatial multiplexing gain, XL-MIMO should generate a directional beam with high array gain by beamforming. Before realizing beamforming, the channel state information (CSI) should be acquired in advance by channel estimation~\cite{GaoZhen21,xueru}. Since the electromagnetic field properties change from far-field to near-field in the XL-MIMO, the existing far-field channel model mismatches the practical near-field XL-MIMO channel feature. In this case, the existing far-field channel estimation methods suffer from serious performance loss in the near-field XL-MIMO channel model. Thus, it is important to delicately model the XL-MIMO channel and design a near-field channel estimation scheme by means of analyzing the propagation property in the near-field XL-MIMO scenario~\cite{Shu}.

\subsection{Prior Contributions}\label{S1.1}
The existing near-field channel estimation methods for ELAA based communication systems can be divided into two categories, i.e., MISO channel estimation~\cite{JinShi,Cui_Tcom} and MIMO channel estimation. For the near-field MISO scenario, the receiver with a single antenna is in the near-field region of the transmitter with ELAA. Considering the near-field effects, the XL-MISO channel should be modeled under the condition of the spherical wave assumption instead of the planar wave assumption as far-field to ensure accuracy in the near-field range~\cite{Yin17,Cui_Tcom}. For example,~\cite{JinShi} considers the scenario where the transmitter employs an ELAA serving multiple single-antenna receivers. In this scenario, the receivers are located in the transmitter's near-field region. Thus, the near-field MISO channel between the transmitter and receiver is modeled based on a near-field array response vectors accurately, which relates not only to the angle but also to the distance due to the spherical wave assumption. To estimate the near-field channel, the whole two-dimensional distance-angle plane is uniformly partitioned into multiple grids, and then the corresponding near-field transform matrix can be constructed by multiple near-field array response vectors associated with different grids. By means of the constructed transform matrix, the near-field channel shows sparsity in the transform domain, which can be estimated by compressive sensing (CS)-based methods with low pilot overhead. The authors of \cite{Cui_Tcom} also consider the same scenario as that in~\cite{JinShi}, interestingly, it is proved that the distance should be non-uniformly divided to reduce the correlation among the near-field array response vectors of the transform matrix in~\cite{Cui_Tcom}. Based on the improved transform matrix, a new sparse representation in polar-domain is proposed in \cite{Cui_Tcom} for XL-MIMO. By utilizing this polar-domain sparsity, a CS-based algorithm has been proposed to increase the estimation accuracy.

\begin{figure}[tp]
	\begin{center}
		\vspace*{-1mm}\hspace*{-1mm}\includegraphics[width=1\linewidth]{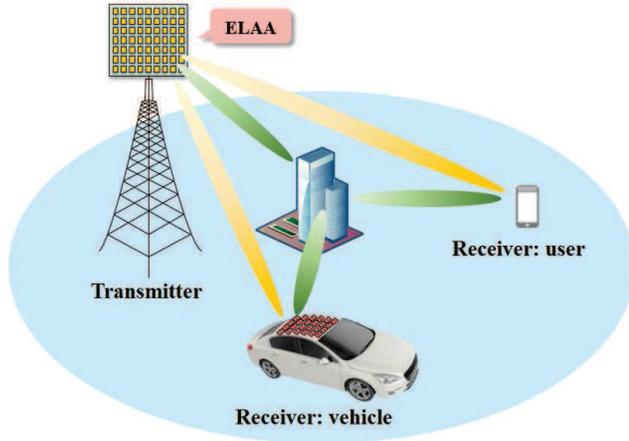}
	\end{center}
	\vspace*{-2mm}\caption{{\color{black}The XL-MIMO communication system with ELAA at the transmitter, where the receiver is a multi-antenna user or a vehicle equipped with an ELAA.}}\vspace{+0mm} \label{scenario}
\end{figure}

The second category is based on the near-field XL-MIMO scenario, where the transmitter employs an ELAA that serves receivers with multiple antennas or even an ELAA as shown in Fig.~\ref{scenario}.  
For instance, an ELAA is installed on the top of a train~\cite{ZIjun,Railway}, and the transmitter also deploys an ELAA. Additionally, ELAA can also be integrated into large infrastructures as revisers, such as the roof of airports and the walls of stadiums. 
{\color{black}Additionally, XL-MIMO-based wireless backhaul for the ultra-dense network (UDN) is also a promising use case in the future 6G~\cite{UDN_6G}. Specifically, in UDN, macrocell base stations (BSs) and many ultra-dense small-cell BSs equipped with ELAAs cooperate to provide a high data rate for users~\cite{UDN_GAO}. A premise to deploy UDN is the reliable and large bandwidth backhaul connecting ultra-dense small-cell BSs and macrocell BSs. In this case, the signal propagation between these two kinds BSs should be modeled under near-field assumption instead of far-field assumption. Therefore, the backhaul connecting in UDN is also a prospective application for near-field XL-MIMO communications.}
There are two typical kinds of methods for modeling near-field XL-MIMO channel, i.e., utilizing approximation by the far-field channel model~\cite{yuhang} and the product of the near-field array response vectors similar to far-field MIMO model~\cite{zhou15,Raj17,Liana21}.
For the first kind, in~\cite{yuhang}, the ELAA at transmitter and receiver are separated into subarrays with a small aperture. The small aperture of the subarray leads to a small near-field region. Thus, the subchannels between these subarrays can be regarded as far-field channels due to the small apertures. In this case, the whole channel between the transmitter and receiver can be approximated by these far-field subchannels. However, the accuracy of this method highly relies on the number of subarrays and the distance between the transmitter and receiver. As for the second kind of method based on the near-field array response vectors, the authors of~\cite{zhou15} draw on the experience of the far-field MIMO channel model and construct the near-field MIMO channel as the product of the transmitter and receiver near-field array response vectors. By utilizing the sparsity in polar-domain, the CS-based method can be applied to the channel estimation problem under this channel model. 

However, the existing near-field XL-MIMO channel model can only accurately describe the non-line-of-sight (NLoS) path component of the near-field MIMO while leading to an inaccurate description of the line-of-sight (LoS) path component. The reasons are explained as below. 1) For the NLoS path component, the signal radiates from the transmitter to the scatterers and then to the receiver. Since the scatterers can be regarded as single-antenna transceivers, the signal actually propagates through two MISO channels. Therefore, the NLoS path component can be presented as the product of transmitter and receiver array response vectors. This modeling method can apply to both far-field and near-field NLoS components. 2) As for the LoS path component, in the far-field scenario, the LoS path component is modeled in the same way as the NLoS path component. This is because the transmitter with a small-aperture antenna array can be viewed as a point from the receiver and vice versa, where the transmitting and receiving processes are equivalent to two MISO propagation processes. Thus, in the far-field scenario, the LoS path component can also be represented by the product of transmitter and receiver array response vectors.
{\color{black}However, in a near-field XL-MIMO system, due to the large aperture of ELAA, the receiver cannot be viewed as a point from the transmitter and vice versa, and the transmitting and receiving processes cannot be approximated by the previous theoretical model of two MISO propagation processes. 
	In this case, there exists a huge gap, which is defined as the accuracy loss, between the propagation distance of the practical LoS path component and that of the theoretical model based on two MISO processes.}
Intuitively, the accuracy loss will increase as the array aperture increases, especially when the transmitter and receiver employ the ELAAs simultaneously. 

In this case, the existing near-field XL-MIMO channel model based on near-field array response vectors mismatches the XL-MIMO scenario in practice. Therefore, the existing near-field channel estimation schemes cannot estimate the XL-MIMO channel accurately. Unfortunately, to the best of our knowledge, there is no study of this problem for XL-MIMO in the current literature.

\subsection{Our Contributions}
In order to solve this problem, an accurate mixed LoS/NLoS near-field XL-MIMO channel model is first proposed. Then, a two stage XL-MIMO channel estimation scheme is proposed for the mixed LoS/NLoS near-field XL-MIMO channel model\footnote{Simulation codes are provided to reproduce the results in this paper: \url{http://oa.ee.tsinghua.edu.cn/dailinglong/publications/publications.html}.}. Our specific contributions are listed below.  

\begin{enumerate}
\item In order to accurately describe the near-field XL-MIMO channel, a mixed LoS/NLoS near-field XL-MIMO channel model is proposed. Specifically, the LoS path components and NLoS path components are modeled separately. The NLoS path components are modeled based on the near-field array response vectors, which is similar to the far-field array response vectors based on the far-field channel model. By contrast, the LoS path component is modeled based on the precise geometric free space assumption.

\item Based on the proposed mixed LoS/NLoS near-field XL-MIMO channel model, we derive the MIMO Rayleigh distance (MIMO-RD) and MIMO advanced Rayleigh distance (MIMO-ARD) to provide the near-field region for the XL-MIMO scenario. 
{\color{black}In specific, similar to the derivation of MISO-RD, by considering the condition that the maximum phase discrepancy between the proposed channel model and the far-field array response vectors based XL-MIMO channel model in the free space cannot exceed $ \pi/8 $, we first define the MIMO-RD, which is proportional to the square of the sum of the antenna array apertures of the transmitter and receiver.} Then, by considering the condition that the maximum phase discrepancy between the proposed channel model and the existing near-field channel model in the free space is no more than $ \pi/8 $, the MIMO-ARD is calculated, which is proportional to the product of the transmitter and receiver array apertures. The derived MIMO-RD and MIMO-ARD show the bound of the proposed channel model between the far-field array response vectors based channel model and near-field array response vectors based channel model, separately.

\item We propose a two stage channel estimation algorithm for the proposed channel model to estimate the XL-MIMO channel accurately. In our proposed two stage channel estimation algorithm, the LoS and NLoS path components are estimated separately. The LoS path component estimation is first realized by searching collection with coarse on-grid parameters, and then refined by iteration optimization, while the NLoS path components are estimated by orthogonal matching pursuit (OMP)-based estimation with their polar-domain sparsity. Moreover, the Cram\'er-Rao lower bound (CRLB) and the complexity of the proposed channel estimation algorithm are derived in this paper. {\color{black}Finally, we provide numerical simulation results to illustrate the effectiveness of our scheme in both the proposed channel model and the QuaDRiGa channel emulation platform.}
\end{enumerate}

\subsection{Organization and Notation}
{\it Organization}: The rest of the paper is organized as follows. In Section \ref{S2}, we first introduce the signal model, and then review existing near-field XL-MISO and XL-MIMO channel models. In Section \ref{S3}, we propose the mixed LoS/NLoS near-field XL-MIMO channel model. {\color{black}Then, the MIMO-RD and MIMO-ARD are derived to determine the range of the XL-MIMO near-field region in Section IV.} In Section \ref{S5}, we propose the corresponding channel estimation scheme based on the proposed channel model, and analyze CRLB and its complexity. Simulation results is provided in Section \ref{S6}, and finally, conclusions are provided in Section \ref{S7}.

{\it Notation}: Lower-case and upper-case boldface letters ${\bf{a}}$ and ${\bf{A}}$ denote a vector and a matrix, respectively; $ ||{\bf{A}}||_F $ denotes the Frobenius norm; ${{{\bf{a}}^H}}$ and ${{{\bf{A}}^{H}}}$ denote the conjugate transpose of vector $\bf{a}$ and matrix $\bf{A}$, respectively. The circularly symmetric complex Gaussian distribution is denotes by ${\cal CN}\left(\mu,\sigma^2 \right)$, with $\mu$ as mean set and $\sigma^2$ as variance, and ${{\cal U}(-a,a)}$ denotes the uniform distribution on $(-a,a)$. $ \otimes $ denotes the Kronecker product. $\bm I$ denotes the identity matrix. $ \rm floor $ denotes round down operation.

\section{System Model}\label{S2}
In this section, the signal model of the near-field XL-MIMO system used in this paper will be introduced first. Then, we will review the existing near-field XL-MISO and XL-MIMO channel models.

\subsection{Signal Model}\label{S2.1}
In this work, we consider that the transmitter and receiver are equipped with $N_1$-element and $N_2$-element antenna arrays, respectively. Since the antenna arrays are usually implemented in a digital-analog hybrid manner with a few RF (radio frequency) chains, we assume that RF chain numbers of the transmitter and receiver are $ {N_{\rm t}^{\rm RF}} $ and $ {N_{\rm r}^{\rm RF}} $. Let ${\bf{H}}\in\mathbb{C}^{{N_2}\times {N_1}}$ denotes the channel from transmitter to receiver. 
The corresponding signal model can be presented as
\begin{equation}\label{eq1.1}
{\bf{y}}_m = {\bf{W}}{\bf{H}}{\bf{Q}}{\bf{s}}_m + {\bf{n}}_m,
\end{equation}
where ${\bf{y}}_m\in\mathbb{C}^{{N_{\rm r}^{\rm RF}}\times 1}$, $ {\bf{W}} \in\mathbb{C}^{{N_{\rm r}^{\rm RF}}\times {N_2}} $, ${{\bf{Q}}}\in\mathbb{C}^{{N_1}\times {N_{\rm t}^{\rm RF}}}$ and $ {\bf{s}}_m \in\mathbb{C}^{{N_{\rm t}^{\rm RF}}\times1 } $ denote the received pilots signal, combining matrix, the hybrid precoding matrix, and transmitted signal in $m$-th times slots, and ${{\bf{n}}_m}\sim{\cal C}{\cal N}\left( {{\bf{0}},\sigma^2{\bf{I}}_{N_{\rm r}^{\rm RF}}} \right)$ denotes the ${{N_{\rm r}^{\rm RF}}\times 1}$ received noise with ${\sigma^2}$ representing the noise power after combining in the $m$-th times slots. 

Denote $ {\bf p}_m = {\bf{Q}}{\bf{s}}_m \in\mathbb{C}^{{N_1}\times 1} $, where the $ i $-th element of the $ {\bf p}_m $ is the signal transmitted by the $ i $-th antenna at transmitter in $m$-th time slots. By collecting the received pilots in $ M $ time slots, we have
\begin{equation}\label{eq1}
{\bf{Y}} = {\bf{W}}{\bf{H}}{\bf{P}} + {\bf{N}},
\end{equation}
where $ {\bf Y} = \left[ {\bf{y}}_1,{\bf{y}}_2,\ldots,{\bf{y}}_M \right]  $, and $ {\bf P} = \left[ {\bf{p}}_1,{\bf{p}}_2,\ldots,{\bf{p}}_M \right]  $ and $ {\bf N} = \left[ {\bf{n}}_1,{\bf{n}}_2,\ldots,{\bf{n}}_M \right]  $.
${{\bf{Y}}}\in\mathbb{C}^{{N_{\rm r}^{\rm RF}}\times M}$, $ {\bf{W}} \in\mathbb{C}^{{N_{\rm r}^{\rm RF}}\times {N_2}} $ and ${{\bf{P}}}\in\mathbb{C}^{{N_1}\times M}$ denote the received pilots signal, combining matrix and the transmitted pilots signal in $M$ times slots in a coherence interval, and ${{\bf{N}}}\sim{\cal C}{\cal N}\left( {{\bf{0}},\sigma^2{\bf{I}}_{{N_{\rm r}^{\rm RF}}} \otimes {\bf{I}}_M} \right)$ denotes the ${{N_2} \times M}$ received noise with ${\sigma^2}$ representing the noise power in $M$ times slots. 
In channel estimation problem, we need to estimate $\bf{H}$ with given ${\bf{P}}$, ${\bf{W}}$ and ${\bf{Y}}$. The number of transmitter $N_1$ is usually large in the XL-MIMO system. Thus, to reduce the pilot overhead in a practical communication system, the channel estimation scheme with low overhead should be utilized $ \left( M < N_1\right) $. 
Since the channel model is significant for designing a channel estimation scheme, we will review the current XL-MIMO channel models next.

\subsection{Existing Near-Field XL-MISO Channel Model}
Most of the existing near-field channel estimation works consider that only the transmitter employs the ELAA, while the receivers are usually equipped with a single antenna, i.e., near-field XL-MISO scenario. The near-field XL-MISO channel between the receiver and the transmitter $ {\bf{h}}_{\rm{n\mbox{-}f}} $ can be represented as
\begin{equation}\label{eq6}
{\bf{h}}_{\rm{n\mbox{-}f}}=\sqrt{\frac{N_1}{L}}\sum\limits_{l = 1}^{L}{\alpha_{l}}{\bf{b}}\left({\theta}_l,r_l\right).
\end{equation}
The near-field array response vector ${\bf{b}}\left({\theta}_l,r_l\right)$ in~(\ref{eq6}) is derived on the base of spherical wave assumption instead of planar wave assumption, which can presented as~\cite{Cui_Tcom}
\begin{equation}\label{eq7}
{\bf{b}}(\theta_l, r_l) = \frac{1}{\sqrt{N_1}}[e^{-j{\frac{2\pi}{\lambda}}(r_{l}^{(1)} - r_{l})},\cdots, e^{-j{\frac{2\pi}{\lambda}}(r_{l}^{(N_1)} - r_{l})}]^H,
\end{equation}
where $r_l$ denotes the distance of the $l$-th scatterer from the center of the transmitter antenna array, $r_{l}^{({n_1})} = \sqrt{r_l^2 + \delta_{n_1}^2d^2 - 2r_l\delta_{n_1} d\sin\theta_l}$ represents the distance of the $l$-th scatterer from the $n_1$-th transmitter antenna, $\delta_{n_1} = \frac{2{n_1} - {N_1} - 1}{2}$ with ${n_1} = 1,2,\cdots, {N_1}$, $\theta_{l}\in(-\pi/2,\pi/2)$ are the practical physical angles, $ d $ is the antenna spacing of half wavelength to avoid coupling between the antennas. {\color{black}It is worth pointing out that a compact array with below half a wavelength spacing can occur in XL-MIMO, which will cause coupling between the antennas. In this case, the coupling effect should be measured first and then modeled as a coupling coefficient matrix with tunable loads~\cite{coupling} in the channel model.}

By utilizing the Taylor expansion $ \sqrt{(1+x)} \approx 1+\frac{1}{2} x-\frac{1}{8}  x^2+\mathcal{O}(x^2) $, the distance difference $r_{l}^{({n_1})} - r_{l}$ in (\ref{eq7}) can be approximated by:
\begin{equation}\label{eq2.1}
r_{l}^{({n_1})} - r_{l} \approx = -\delta_{n_1}d{\sin\theta_l}+\delta_{n_1}^2d^2\frac{{\cos^2\theta_l} }{2{r_{l}}}.
\end{equation}

{\color{black}To exploit the sparsity of near-field channel,~\cite{Cui_Tcom} proposed a new transform matrix $\bf{D}$ to change the channel $ {{\bf{h}}_{{\rm{n\mbox{-}f}}}} $ in~(\ref{eq6}) into polar-domain, which can be presented as
\begin{equation}\label{eq8}
\begin{aligned}
{\bf{D}}=[{\bf{b}}(\theta_1, r_1^1),\cdots,{\bf{b}}(\theta_1, r_{1}^{S_{1}}),\cdots,\\{\bf{b}}(\theta_{N_1}, r_{N_1}^1),\cdots,{\bf{b}}(\theta_{N_1}, r_{N_1}^{S_{N_1}})],
\end{aligned}
\end{equation}
where each column of polar-domain transform matrix $\bf{D}$ is a near-field array response vector sampled on the grid  ($\theta_{n_1}$, $r_{{n_1}}^{s_{n_1}}$), 
with $s_{n_1}=1,2,\cdots,{S_{n_1}}$, $S_{n_1}$ denotes the number of sampled distances at $\theta_{{n_1}}$. Therefore, the number of total sampled grids of the whole propagation environment is $S=\sum\limits_{{n_1} = 1}^{N_1}S_{n_1}$. With help of $\bf{D}$,  ${\bf{h}}_{\rm{n\mbox{-}f}}$ in (\ref{eq6}) can be presented as
\begin{equation}\label{eq9.2}
{\bf{h}}_{\rm{n\mbox{-}f}}= {\bf{D}}{{\bf{h}}_{\rm{n\mbox{-}f}}^P},
\end{equation}
where ${{\bf{h}}_{\rm{n\mbox{-}f}}^P}$ is the channel represented in polar-domain of size $S\times 1$.} This near-field ${{\bf{h}}_{\rm{n\mbox{-}f}}^P}$ in (\ref{eq9.2}) shows certain sparsity in the polar-domain. The authors in~\cite{Cui_Tcom} proposed the associated polar-domain based CS algorithm to solve the near-field channel estimation problem.

\subsection{Existing Near-Field XL-MIMO Channel Models}\label{S2.2}
{\color{black}The existing near-field XL-MIMO channel model refers to the far-field MIMO channel model. Specifically, the far-field channel model is based on the production of the far-field array response vectors. The near-field array response vectors are utilized to replace the far-field array response vectors to model near-field XL-MIMO channel~\cite{zhou15}, which is presented as
\begin{equation}\label{eq3.1}
{\bf{H}}_{\rm{n\mbox{-}f}} = \sum\limits_{l = 1}^L {{g_l} {\bf{b}} (\theta _r^l,d _r^l)} { {\bf{b}} ^H}(\theta _t^l,d _t^l),
\end{equation}
where $ L $ denotes the number of path components, $ {g_l} $ represents the complex gain.} ${\bf{ b}} (\theta _t^l,d _t^l) $ and $  {\bf{b}} (\theta _r^l,d _r^l) $ denote the near-field array response vectors at transmitter and receiver on the base of spherical wave assumption, which is denoted by
\begin{equation}\label{eq7.1}
\begin{aligned}
{\bf{b}}(\theta _t^l, d_t^l) = \frac{1}{\sqrt{N_1}}[e^{-j{\frac{2\pi}{\lambda}}(d_{t}^{l}(1) - d_t^l)},\cdots, e^{-j{\frac{2\pi}{\lambda}}(d_{t}^{l}(N_1) - d_t^l)}]^H, \\
{\bf{b}}(\theta _r^l, d_r^l) = \frac{1}{\sqrt{N_2}}[e^{-j{\frac{2\pi}{\lambda}}(d_{r}^{l}(1) - d_r^l)},\cdots, e^{-j{\frac{2\pi}{\lambda}}(d_{r}^{l}(N_2) - d_r^l)}]^H,
\end{aligned}
\end{equation}
where $ \theta _t^l\ (\theta _r^l) $ represent the angle for the $ l $-th path at transmitter (receiver), and $ d _t^l\ (d _r^l) $ represent the distance of the $ l $-th scatterer from the center of the antenna array of transmitter (receiver) for the $ l $-th path,  
$d_{t}^{l}({n_1}) = \sqrt{{d_t^l}^2 + \delta_{n_1}^2d^2 - 2{d_t^l}\delta_{n_1} d{\sin\theta _t^l}}$ 
represents the distance of the $l$-th scatterer from the $n_1$-th element on transmitter antenna array, and $\delta_{n_1} = \frac{2{n_1} - N_1 - 1}{2}$ with $n = 1,2,\cdots, N_1$, and $d_{r}^{l}(n_2) = \sqrt{{d_r^l}^2 + \delta_{n_2}^2d^2 - 2{d_r^l}\delta_{n_2} d{\sin\theta _r^l}}$ 
represents the distance of the $l$-th scatterer from the $n_2$-th transmitter antenna array, and $\delta_{n_2} = \frac{2{n_2} - N_2 - 1}{2}$ with $n = 1,2,\cdots, N_2$.


The polar-domain transform matrix mentioned in (\ref{eq8}) can be presented as
\begin{equation}\label{eq8.1}
\begin{aligned}
{\bf{D}}_t=[{\bf{b}}(\theta_{1}, d_1^1),\cdots,{\bf{b}}(\theta_1, d_{1}^{S_{1}}),\cdots,\\ {\bf{b}}(\theta_{N_1}, d_{N_1}^1),\cdots, {\bf{b}}(\theta_{N_1}, r_{N_1}^{S_{N_1}})],\\
{\bf{D}}_r=[{\bf{b}}(\theta_1, r_1^1),\cdots,{\bf{b}}(\theta_1, d_{1}^{S_{1}}),\cdots,\\ {\bf{b}}(\theta_{{N_2}}, r_{{N_2}}^1),\cdots,{\bf{b}}(\theta_{{N_2}}, r_{{N_2}}^{S_{N_2}})],
\end{aligned}
\end{equation}
where each column of the matrix ${\bf{D}}_t$ (${\bf{D}}_r$) is a near-field array response vector sampled at angle $\theta_{n_1}(\theta_{n_2})$ and distance $d_{n_1}^{s_{n_1}}(d_{n_2}^{s_{n_2}})$, with $s_n=1,2,\cdots,{S_{n_1}}({S_{n_2}})$. ${S_{n_1}}({S_{n_2}})$ denotes the number of sampled distances at the sampled angle $\theta_{n_1}(\theta_{n_2})$. Therefore, we can calculate the total number of all sampled grids, i.e., the number of ${\bf{D}}_t$ (${\bf{D}}_r$) columns, which can be presented as ${S_1}=\sum\limits_{{n_1} = 1}^{N}S_{n_1}({S_2}=\sum\limits_{{n_2} = 1}^{N}S_{n_2})$.

\vspace{2mm}
On the base of this polar-domain transform matrix ${\bf{D}}_t$ and ${\bf{D}}_r$, the channel $ {{\bf{H}}}_{\rm{n\mbox{-}f}} $ can be represented by
\begin{equation}\label{eq9.1}
{{\bf{H}}}_{\rm{n\mbox{-}f}} = {\bf{D}}_r{{\bf{H}}}^P_{\rm{n\mbox{-}f}}{\bf{D}}_t^H,
\end{equation}
where ${{\bf{H}}}^P_{\rm{n\mbox{-}f}}$ is the $S_2 \times S_1$ polar-domain XL-MIMO channel, which also shows sparsity in the polar-domain.

In the existing near-field XL-MIMO channel model above, the LoS path component is modeled as the same method for NLoS path components, 
where the MIMO channel in (\ref{eq3.1}) can be modeled based on two MISO channels. This method is not suitable for the LoS path component in a near-field XL-MIMO scenario. 
Therefore, the existing near-field channel model will mismatch the LoS path component of the near-field XL-MIMO.
Thus, the existing near-field channel estimation algorithms cannot be utilized to address the XL-MIMO channel estimation problem accurately. In order to design the channel estimation algorithm for the XL-MIMO, we should first provide an accurate description of the LoS path component, which will be described in Section \ref{S3}.



\vspace{0mm}
\section{The Proposed Mixed LoS/NLoS Near-Field XL-MIMO Channel Model}\label{S3}
In this section, we will first 
utilize the free space propagation assumption to accurately model the LoS path component for each transmitter-receiver antenna pair. Then, the mixed LoS/NLoS near-field XL-MIMO channel model is provided to capture the different features of LoS and NLoS path components, which are modeled separately.
\begin{figure}[tp]
	\begin{center}
		\vspace*{0mm}\hspace*{0mm}\includegraphics[width=0.92\linewidth]{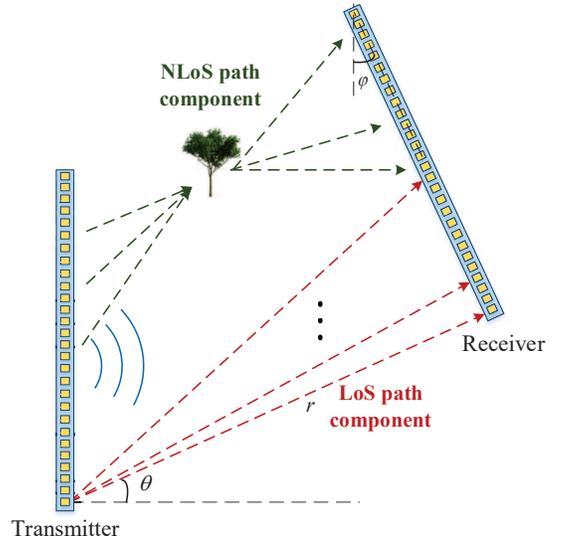}
	\end{center}
	\vspace*{-4mm}\caption{{\color{black}The proposed mixed LoS/NLoS near-field channel model for XL-MIMO.}}\vspace{1mm} \label{FIG1}
\end{figure}

\subsection{LoS Path Component} 
For the near-field XL-MIMO LoS path component, each transmitter-receiver antenna pair will experience different propagation paths as shown in Fig.~\ref{FIG1}.
Thus, in this case, the channel model built with the near-field array response vectors mismatches the practical feature of XL-MIMO near-field LoS path component. As a result, instead of utilizing near-field array response vectors like NLoS path components, we model the LoS path component under the geometric free space propagation assumption~\cite{RAL} for each transmitter-receiver antenna pair. Specifically, $ {{\bf{H}}}({n_2},{n_1}) $ denotes the LoS path component of channel between the transmitter's $ n_1 $-th antenna and the receiver's $ n_2 $-th antenna, which can be represented as 
 \begin{equation}\label{hn2n1}
 \begin{aligned}
 {{\bf{H}}}({n_2},{n_1}) &= {{\frac{1}{{r_{{n_2},{n_1}}}}}e^{ - j\frac{2\pi}{\lambda} {r_{{n_2},{n_1}}}}},
 \end{aligned}
 \end{equation}
where $ {{r_{n_2,n_1}}} $ denotes the distance of the $ n_1 $-th antenna at receiver from the $ n_2 $-th antenna at transmitter. {\color{black}It is worth to point out that the free space path loss of each transmitter-receiver antenna pair has been normalized as $ {\frac{1}{{r_{{n_2},{n_1}}}}} $ in this paper}. {\color{black}The $ {{r_{n_2,n_1}}} $ can be represented as
\begin{equation}\label{Rn1n2}
\begin{aligned}
{r_{n_2,n_1}} &= \sqrt {(r\cos\theta-d_2\sin\phi)^2+(r\sin\theta+d_2\cos\phi-d_1)^2}\\
&= \sqrt {{r^2}\! +\! d_1^2 \!+\! d_2^2\! +\! 2\!\left( r{d_2}\sin ( \varphi\!+\!\theta  )\!\! -\! \! r{d_1}\sin \theta \!\! -\!\! {d_1}{d_2}\cos \varphi\right)  } 
\end{aligned}
\end{equation}
where the $ r $ is the distance of the 1-st antenna at receiver from the 1-st antenna at transmitter, $ \varphi $ denotes relative angle between receiver and transmitter, and $ \theta $ denotes the angle of departure (AoD) of the signal. {\color{black}Moreover, $ d_1 $ and $d_2 $ are denoted as $d_1 = n_1d $ and $ d_2 = n_2d $, where $ d $ is antenna spacing.}} Thus, by utilizing geometry relation in free space, the channel can be presented as
\begin{equation}\label{eq3.5}
{{\bf{H}}_{{\rm{LoS}}}} = \mathbf{H}_{\mathrm{LoS}}(r, \theta, \varphi)={\left[ {{\frac{1}{{r_{{n_2},{n_1}}}}}{e^{ - j2\pi {r_{{n_2},{n_1}}}/\lambda }}} \right]_{{N_2} \times {N_1}}},
\end{equation}

Unlike the NLoS path components, the LoS path component cannot be decoupled by near-field array response vectors. Thus, the LoS path component cannot be presented by polar-domain channel with transform matrices.

\subsection{Proposed Mixed LoS/NLoS Near-Field XL-MIMO Channel}
In order to capture both LoS and NLoS path components features, we propose a mixed LoS/NLoS near-field XL-MIMO channel model based on (\ref{eq9.1}) and (\ref{eq3.5}). The proposed XL-MIMO channel model can be presented:

\begin{equation}
\begin{aligned}
{\bf{H}} & = {\bf{H}}_{{\rm{LoS}}}+{\bf{H}}_{{\rm{NLoS}}}\\
& = \mathbf{H}_{\mathrm{LoS}}(r, \theta, \varphi)+{\bf{D}}_r\left( \sum_{l=1}^{L} g_{l} \mathbf{b}\left(\theta_{r}^{l}, d_{r}^{l}\right) \mathbf{b}^{H}\left(\theta_{t}^{l}, d_{t}^{l}\right)\right) {\bf{D}}_t^H.
\end{aligned}
\end{equation}

\vspace{2mm}

It is worth noting that the LoS path component will degenerate into the same representation as NLoS path components 
as the distance of the transmitter from the receiver increases to a certain extent. Furthermore, if the distance continues to increase, the proposed mixed LoS/NLoS near-field XL-MIMO channel model will degenerate into a far-field MIMO channel model. To prove this statement, in the next Section IV, 
we will derive the boundary between the proposed mixed LoS/NLoS near-field XL-MIMO channel model and far-field MIMO channel model and define the Rayleigh distance for the XL-MIMO scenario. Then, the boundary between the proposed mixed LoS/NLoS near-field XL-MIMO and the existing near-field XL-MIMO is provided and the advanced Rayleigh distance for the XL-MIMO scenario is defined.

\vspace{+0mm}
\section{The Definitions of Rayleigh Distance for Near-Field XL-MIMO}\label{S4}

{\color{black}In this section, we will first define the MIMO Rayleigh distance (MIMO-RD) to determine the boundary between the proposed mixed LoS/NLoS near-field XL-MIMO channel model and the far-field MIMO channel model.} Then, MIMO advanced Rayleigh distance (MIMO-ARD) is derived to determine the boundary between the proposed  mixed LoS/NLoS near-field XL-MIMO and the existing near-field XL-MIMO channel model.

As described in \cite{Ray}, the RD for a MISO scenario, i.e., MISO-RD, is defined as $ Z = 2D^2/\lambda $, where $ D $ is the aperture of the antenna array and $ \lambda  $ is the wavelength. MISO-RD is calculated by the condition that the largest phase discrepancy between the far-field planar wavefront and the near-field spherical wavefront in the free space is no more than $ \pi/8 $. 
{\color{black}However, the current MISO-RD is identified based on the scenario where only the transmitter employs the ELAA while the receiver is equipped with a single antenna. In this case, the MISO-RD is only related to the aperture of the transmitter ELAA. Thus, MISO-RD is not suitable for the XL-MIMO scenario. }
In this paper, by considering the XL-MIMO scenario, MIMO Rayleigh distance (MIMO-RD) is defined by the condition that the largest phase discrepancy between the far-field planar wavefronts and the near-field spherical wavefronts is no more than $ \pi/8 $. 
Similar to MISO-RD, the proposed MIMO-RD describes the boundary between the far-field and near-field regions for the XL-MIMO scenario. However, MIMO-RD is unable to capture the feature of the largest phase discrepancy between the proposed channel model and the existing near-field array response vectors based channel model. 
Thus, we further define the MIMO-ARD by the condition that the largest phase discrepancy between the proposed mixed LoS/NLoS near-field XL-MIMO and the existing near-field MIMO scenario in the free space is no more than $ \pi/8 $. The MIMO-RD and MIMO-RD are derived as below. 

\begin{figure}[tp]
	\begin{center}
		\vspace*{0mm}\hspace*{0mm}\includegraphics[width=0.9\linewidth]{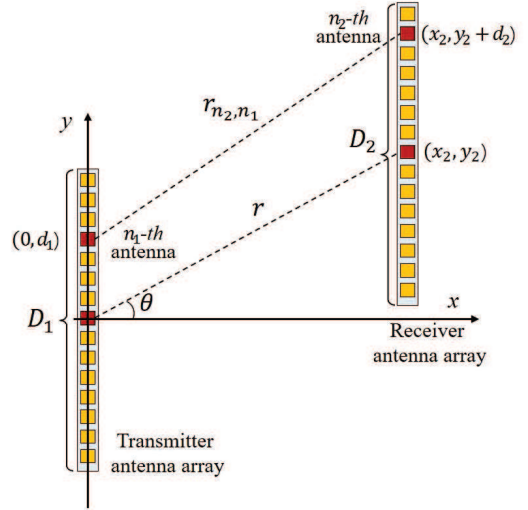}
	\end{center}
	\vspace*{-4mm}\caption{The near-field MIMO scenario: transmitter and receiver are both equipped with ELAAs with antenna array elements (orange squares) in the {\it{x-y}} coordinate system}\vspace{0mm} \label{EDSRD}
\end{figure}

As shown in Fig. \ref{EDSRD}, we consider the scenario where the transmitter and receiver are both equipped with antenna arrays of aperture $ D_1 $ and $ D_2 $, respectively. These two antenna arrays are set in parallel since the largest phase
discrepancy occurs when the wave impinges perpendicularly~\cite{emil2022}. The
center of transmitter antenna arrays is regarded as the $ x-y $ coordinate origin. The coordinate of the $ n_1 $-th antenna of the transmitter antenna array, the center of receiver antenna array, the $ n_2 $-th antenna of the receiver antenna array are $ (0,d_1) $, $ (x_2,y_2) $, and $ (x_2,y_2+d_2) $, respectively, where $ -\frac{D_1}{2} \leq d_1 \leq \frac{D_1}{2} $ and  $ -\frac{D_2}{2} \leq d_2 \leq \frac{D_2}{2} $. 
The polar coordinate of the center of UE's antenna array can be written as $(r, \theta) = \left(\sqrt{x_2^2 + y_2^2}, \arctan(\frac{y_2}{x_2})\right)$, where $ r $, $\theta$ are the distance and angle between the center of transmitter's antenna array and the center of receiver's antenna array, respectively.

\begin{figure*}[tbhp]
	\begin{center}
		\vspace*{0mm}\hspace*{0mm}\includegraphics[width=0.75\linewidth]{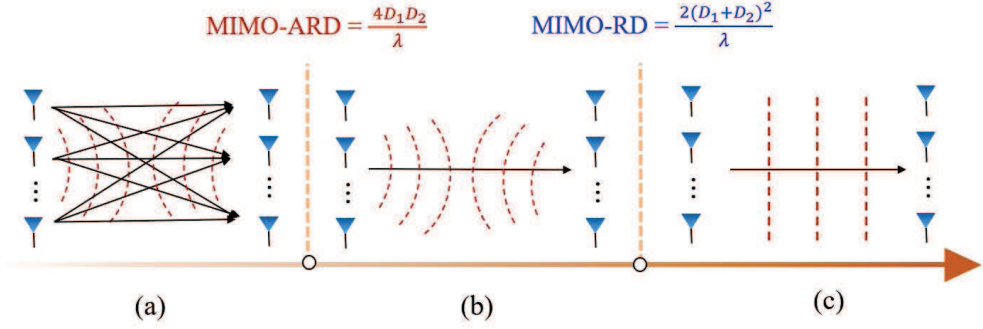}
	\end{center}
	\vspace*{-4mm}\caption{\color{black}Three categories of XL-MIMO divided by the distance of the transmitter from the receiver. 
	}\vspace{-1mm} \label{DSCLASS}
\end{figure*}

To derive the close form of MIMO-RD and MIMO-ARD, we only consider the phase change caused by LoS path component of the channel. 
As mentioned in (\ref{hn2n1}), 
$ {{\bf{H}}}({n_2},{n_1}) = {{\frac{1}{{r_{{n_2},{n_1}}}}}{e^{ - j2\pi {r_{{n_2},{n_1}}}/\lambda }}}  $ represents the LoS path component of channel between the transmitter's $ n_1 $-th antenna and the receiver's $ n_2 $-th antenna.
$ {r_{{n_2},{n_1}}} $ is the distance of the transmitter's $ n_1 $-th antenna from the receiver's $ n_2 $-th antenna. Here, the true phase is $\phi = \frac{2\pi}{\lambda}r_{{n_2},{n_1}}$, where $ {r_{{n_2},{n_1}}} $ is presented as 
\begin{equation}\label{R3.eq9}
\begin{aligned}
{r_{{n_2},{n_1}}} &= \sqrt{x_2^2 + (y_2 + d_2 - d_1)^2} \\
&=r\sqrt{1+\frac{({d_2}-{d_1})^2}{r^2}+\frac{2({d_2}-{d_1})\sin\theta}{r}},
\end{aligned}
\end{equation}
where $ \theta $ is the practical physical angle of departure. Specifically, by utilizing the second-order Taylor expansion $ \sqrt{1+x} \approx 1+\frac{1}{2} x-\frac{1}{8}  x^2+\mathcal{O}(x^2) $, we have
\begin{equation}\label{R3.eq3.6}
\begin{aligned}
{r_{{n_2},{n_1}}}& = r( 1+\frac{({d_2}-{d_1})^2}{2r^2}+\frac{({d_2}-{d_1})\sin\theta}{r}\\
&-\frac{1}{8}(\frac{2({d_2}-{d_1})\sin\theta}{r})^2 + \mathcal{O}(\frac{1}{r^2}) ) \\
&\approx r + {(d_2 -d_1)}\sin\theta+\frac{(d_2 - d_1)^2\cos^2\theta}{2r}.
\end{aligned}
\end{equation}

\subsection{MIMO Rayleigh Distance (MIMO-RD)}

Based on the far-field assumption, ${r_{{n_2},{n_1}}}$ can be approximated by its first-order Taylor expansion. Since $\sqrt{1+x} \approx 1 + \frac{1}{2}x$, we have 
\begin{equation}\label{eq14}
\begin{aligned}
r^{\text{far}}_{n_2, n_1} &\approx r (1 + \frac{(d_2 - d_1)\sin\theta}{r})
& = r + (d_2 - d_1)\sin\theta
\end{aligned}
\end{equation}
Thus, the far-field phase becomes $\phi^{\text{far}} = \frac{2\pi}{\lambda}r^{\text{far}}_{n_2, n_1}$. 
Accordingly, the phase discrepancy between the far-field planar and near-field spherical wavefronts can be presented as
\begin{equation}\label{eq15}
\Delta =  \left|\phi_{n_2, n_1} -\phi_{n_2, n_1}^{\text{far}}\right| = \frac{2\pi}{\lambda} \left|r_{n_2, n_1} - r_{n_2, n_1}^{\text{far}}\right|
\end{equation}
Notice that $r_{n_2, n_1}^{\text{far}}$ in (\ref{eq14}) is the first-order Taylor expansion of $r_{n_2, n_1}$, so $\left|r_{n_2, n_1} - r_{n_2, n_1}^{\text{far}}\right|$ in (\ref{eq15}) is mainly determined by \emph{the second-order Taylor expansion term of $r_{n_2, n_1}$}. Therefore, the phase discrepancy between planar wavefronts and spherical wavefronts is
\begin{equation}
\begin{aligned}
\Delta \approx \frac{\pi(d_2 - d_1)^2\cos^2\theta}{\lambda r}.
\end{aligned}
\end{equation}

Since $\theta \in [-\frac{\pi}{2}, \frac{\pi}{2}]$, $-\frac{D_1}{2} \le d_1 \le \frac{D_1}{2}$, and $-\frac{D_2}{2} \le d_2 \le \frac{D_2}{2}$, it can be easily observed that when $ \theta = 0 $, $ d_1 = \dfrac{D_1}{2} (-\dfrac{D_1}{2}) $ and $ d_2 = -\dfrac{D_2}{2} (\dfrac{D_2}{2}) $, phase discrepancy achieves maximum, where 
\begin{equation}\label{R3.MPD_MIMO}
\begin{aligned}
{\rm{max}}\frac{\pi(d_2 - d_1)^2\cos^2\theta}{\lambda r} = \frac{\pi({D_1}+{D_2})^2}{4r\lambda}.
\end{aligned}
\end{equation}

Eventually, since the largest phase discrepancy is larger than $ \pi/8 $ in the near-field region,
the distance $r$ should satisfy $\frac{\pi({D_1}+{D_2})^2}{4r\lambda} > \frac{\pi}{8}$, where 
\begin{equation}\label{MIMO_RD}
r\leq\frac{2(D_1+D_2)^2}{\lambda}.
\end{equation}
Thus, the MIMO-RD systems can be defined as $ \frac{2(D_1+D_2)^2}{\lambda} $. 


%

\subsection{MIMO Advanced Rayleigh Distance (MIMO-ARD)}
In the existing near-field XL-MIMO channel model, the LoS and NLoS path components are both modeled based on the product of two near-field response vectors. In this case, the phase of the signal from the $ n_1 $-th antenna of the transmitter to the $ n_2 $-th antenna of the receiver can be acquired by the sum of the phase of the $ n_1 $-th phase of the array response vector at the transmitter and the $ n_2 $-th phase of array response vector at the receiver. As delivered in (\ref{eq2.1}), the phase of the near-field array response vector can be approximated by its second-order Taylor expansion. We utilize this approximation to calculate the sum of the phase of the $ n_1 $-th phase of the transmitter array response vector and the $ n_2 $-th receiver phase of the array response vector. Thus, the phase of the signal from the $ n_1 $-th antenna of the transmitter to the $ n_2 $-th antenna of the receiver can be represented as
\begin{equation}\label{eq3.11}
\begin{aligned}
{\phi_{{n_2},{n_1}}^\text{near}} &\approx 
&=-{d_1}\sin\theta+\frac{\cos^2\theta{d_1}^2}{2r}+{d_2}\sin\theta+\frac{\cos^2\theta{d_2}^2}{2r}.
\end{aligned}
\end{equation}
Accordingly,based on (\ref{R3.eq3.6}) and (\ref{eq3.11}), the phase discrepancy between the existing near-field channel model and true near-field channel model can be presented as
\begin{equation}\label{eq15}
\begin{aligned}
\Delta &=  \left|\phi_{n_2, n_1} -\phi_{n_2, n_1}^{\text{near}}\right|
& \approx \frac{2\pi}{\lambda} \left|\frac{\cos^2\theta{d_1}{d_2}}{r}\right| 
\end{aligned}
\end{equation}

Since $\theta \in [-\frac{\pi}{2}, \frac{\pi}{2}]$, $-\frac{D_1}{2} \le d_1 \le \frac{D_1}{2}$, and $-\frac{D_2}{2} \le d_2 \le \frac{D_2}{2}$, it can be easily observed that when $ \theta = 0 $, $ d_1 = \dfrac{D_1}{2} (-\dfrac{D_1}{2}) $ and $ d_2 = \dfrac{D_2}{2} (- \dfrac{D_2}{2}) $, phase discrepancy achieves maximum, where 
\begin{equation}\label{R3.MPD_MIMO_E}
\begin{aligned}
{\rm{max}}\frac{2\pi{\cos^2\theta{d_1}{d_2}}}{\lambda r} = \frac{\pi{D_1}{D_2}}{2r\lambda}.
\end{aligned}
\end{equation}

Eventually, since the largest phase discrepancy is larger than $ \pi/8 $ in the near-field region,
the distance $r$ should satisfy $\frac{\pi{D_1}{D_2}}{2r\lambda} > \frac{\pi}{8}$, where 
\begin{equation}\label{MIMO_ERD}
r<\frac{4D_1D_2}{\lambda}.
\end{equation}

Thus, the MIMO-ARD can be defined as $ \frac{4D_1D_2}{\lambda} $. 

From the derivation above, the XL-MIMO can be divided into three categories as shown in Fig.~\ref{DSCLASS}, where the LoS path component of the XL-MIMO channel has different features due to the different distances of the transmitter from the receiver. Specifically, when the distance is bigger than the MIMO-RD, the phase discrepancy between the existing near-field scenario and the far-field scenario can be ignored. In this case, the LoS path component can be modeled by the product of far-field array response vectors at the transmitter and receiver. 
Additionally, when the distance of the transmitter from the receiver is bigger than the MIMO-ARD and smaller than the MIMO-RD, the phase discrepancy between the LoS path components of the proposed mixed LoS/NLoS near-field channel and the existing near-field channel can be ignored. In this case, the LoS path component can be modeled by the multiplication of near-field response vectors at transmitter and receiver.
Furthermore, when the distance of the transmitter from the receiver is smaller than the MIMO-ARD, the phase discrepancy between the LoS path components of the proposed mixed LoS/NLoS near-field channel and the existing near-field channel cannot be ignored. 

In this paper, we mainly focus on the scenario where the distance of the transmitter from the receiver is smaller than the MIMO-ARD. In the following Section V, the specific channel estimation scheme will be described.

\vspace{0mm}
\section{Proposed Two Stage Channel Estimation Algorithm}\label{S5}
In this section, based on the proposed mixed LoS/NLoS near-field XL-MIMO channel model, we propose the two stage near-field XL-MIMO channel estimation algorithm, where a parameter estimation algorithm is utilized for the LoS path component estimation and an OMP-based algorithm is utilized for the NLoS path components estimation. At last, we analyze the CRLB and the computational complexity of the proposed two stage channel estimation algorithm.

\subsection{Stage 1: LoS Path Component Estimation}
{\color{black}Since the energy of the LoS path component is usually dominant, 
we will first conduct the LoS path component estimation.} From (\ref{eq3.5}), we can observe that the LoS path component of near-field MIMO channel is determined by three parameters, i.e., the distance of the 1-st antenna at receiver from the 1-st antenna at transmitter $ r $, relative angle between receiver and transmitter $ \varphi $, and the AoD $ \theta $. Therefore, the LoS path component estimation of (\ref{eq1}) can be recognized as a parameter estimation problem, which can be presented as
\begin{equation}\label{eq4.10}
\begin{aligned}
\mathop{\min} \limits_{r, \theta, \varphi} G(r, \theta, \varphi) 
\buildrel \Delta \over 
=& 
{\left\| {{\bf{Y}} - \mathbf{W}\mathbf{H}_{\mathrm{LoS}}(r, \theta, \varphi)\bf{P}} \right\|^2_F}.
\end{aligned}
\end{equation}
\textbf{Algorithm 1} shows the specific procedure to solve the problem (\ref{eq4.10}).

{\begin{algorithm}[htbp]
{	\color{black}\caption{LoS path component estimation}}
	\textbf{Inputs}: Received signal ${\bf{Y}}$, pilot ${\bf{P}}$, $R_{max}$, $R_{min}$, $\theta_{max}$, $\theta_{min}$, $\varphi_{max}$, $\varphi_{min}$,
	$ r_s $, $ \theta_s $, $ \varphi_s $, $ I $, {\color{black}$\epsilon$}. 		
	\\\textbf{Initialization}:  $ \Delta r_1 = \frac{r_{max}-r_{min}}{r_s} $, $ \Delta \theta_1= \frac{\theta_{max}-\theta_{min}}{\theta_s} $, $ \Delta \varphi_1 = \frac{\varphi_{max}-\varphi_{min}}{\varphi_s} $, calculate $ \Xi $ based on (\ref{eq4.2})
	\\ // Estimate coarse on-grid parameters
	\\1. \textbf{for}  $ \left( {r,\theta ,\varphi } \right) \in \Xi $ \textbf{do}
	\\2. \hspace*{+3mm} calculate $ {\bf{H}}_{{\rm{LoS}}}^{*} $ based on (\ref{eq3.5})
	\\3. \hspace*{+3mm} ${r^{{\rm{int}}},\theta^{{\rm{int}}},\varphi^{{\rm{int}}}}={\mathop{\rm{argmin}}\limits_{r,\theta ,\varphi }}\|{\bf{Y}}- {\bf{H}}_{{\rm{LoS}}}{\bf{P}} \|^2_2$
	\\4. {\bf{end for}}
	\\ // Refine off-grid parameters
	\\5. \textbf{for} $ i \in I $ \textbf{do}
	\\6. \hspace*{+3mm}update $ \hat{{r}}^{(i+1)} $ based on (\ref{eq4.9})
	\\7. \hspace*{+3mm}update $ \hat{{\theta}}^{(i+1)} $ based on (\ref{eq4.9.1})
	\\8. \hspace*{+3mm}update $ \hat{{\theta}}^{(i+1)} $ based on (\ref{eq4.9.2})
{	\color{black}\\9. \hspace*{+3mm}\textbf{if} $ \left| \frac{\hat{{r}}^{(i+1)}-\hat{{r}}^{(i)}}{\hat{{r}}^{(i+1)}}  \right|$ $\leq$ $\epsilon$ and $ \left| \frac{\hat{{\theta}}^{(i+1)}-\hat{{\theta}}^{(i)}}{\hat{{\theta}}^{(i+1)}}  \right|$ $\leq$ $\epsilon$ and $ \left| \frac{\hat{{\theta}}^{(i+1)}-\hat{{r}}^{(i)}}{\hat{{\theta}}^{(i+1)}}  \right|$ $\leq$ $\epsilon$  \textbf{then}
\\10.\hspace*{+3mm}\hspace*{+3mm}$  ({r^{\rm{opt}},{\theta}^{\rm{opt}} ,{\varphi}^{\rm{opt}} }) $ = $ ({\hat{{r}}^{(i+1)},{\hat \theta}^{(i+1)},{\hat \varphi}^{(i+1)}}) $
\\11.\hspace*{+3mm}\hspace*{+3mm}\textbf{jump to} Step 15
\\12.\hspace*{+3mm}\textbf{end if }}
	\\13.\textbf{end for }
	\\14.$  ({r^{\rm{opt}},{\theta}^{\rm{opt}} ,{\varphi}^{\rm{opt}} }) $ = $ ({\hat{{r}}^{(I)},{\hat \theta}^{(I)},{\hat \varphi}^{(I)}}) $
	\\15.calculate ${\hat{{\bf{H}}}}_{{\rm{LoS}}}$ based on (\ref{eq3.5}) by $ ( {r^{\rm{opt}},{\theta}^{\rm{opt}} ,{\varphi}^{\rm{opt}} } )$
	\\\textbf{Output}: Estimated LoS path component ${\hat{{\bf{H}}}}_{{\rm{LoS}}}$.
\end{algorithm}}
\vspace{+0mm}

First, in Steps1-4 we obtain on-grid coarse parameters estimates by searching the collection $ \Xi $, which can be presented as
\begin{equation}\label{eq4.2}
\begin{aligned}
\Xi = \{ \left( {r,\theta ,\varphi } \right)  | 
r &= r_{\rm{min}}, r_{\rm{min}}+\Delta r, \cdots, r_{\rm{max}};\\
\theta &= \theta_{\rm{min}}, \theta_{\rm{min}}+\Delta \theta, \cdots, \theta_{\rm{max}};\\
\varphi &= \varphi_{\rm{min}}, \varphi_{\rm{min}}+\Delta \varphi, \cdots, \varphi_{\rm{max}}  \},
\end{aligned}
\end{equation}
where $ r_{\rm{min}} $, $ r_{\rm{max}} $, $ \theta_{\rm{min}} $, $ \theta_{\rm{max}}$, $ \varphi_{\rm{min}} $, and $ \varphi_{\rm{max}} $  represent the lower and upper boundaries of the distance of the 1-st antenna at receiver antenna array from the 1-st antenna at transmitter antenna array $ r $, relative angle between receiver and transmitter $ \varphi $, and the AoD of the signal $ \theta $, respectively. $ \Delta r $, $ \Delta \theta $, and $ \Delta \varphi $ are the step sizes of $ r $, $ \theta $, and $ \varphi $. 

After searching the collection $ \Xi $, we can get $  {r^{\rm{int}},{\theta}^{\rm{int}} ,{\varphi}^{\rm{int}} } $ as the initial on-grid estimated parameters. Then, to obtain the accurate estimated parameters, we refine three parameters $  {r^{\rm{int}},{\theta}^{\rm{int}} ,{\varphi}^{\rm{int}} } $ by iterative optimization method. Specifically, we define the objective function $  G(r, \theta, \varphi) $ as 
\begin{equation}\label{eq4.1}
\begin{aligned}
\mathop{\min} \limits_{r, \theta, \varphi} G(r, \theta, \varphi) 
\buildrel \Delta \over 
=& 
{\left\| {{\bf{Y}} - \mathbf{W}\mathbf{H}_{\mathrm{LoS}}(r, \theta, \varphi)\bf{P}} \right\|^2_F}, \\
=& \sum_{m=1}^{M}\left({\bf{y}}_m-{\bf W}{\bf{H}}{\bf{p}}_m \right) ^H\left({\bf{y}}_m-{\bf W}{\bf{H}}{\bf{p}}_m \right) \\
=& \sum_{m=1}^{M} {\bf{p}}_m^H{\bf{H}}^H{\bf W}^H{\bf W}{\bf{H}}{\bf{p}}_m\!\!+\!\!\sum_{m=1}^{M}{\bf{y}}_m^H{\bf{y}}_m\\&-\!\!\sum_{m=1}^{M}\left( {\bf{p}}_m^H{\bf{H}}^H{\bf W}^H{\bf{y}}_m\!\!+\!\!\left( {\bf{p}}_m^H{\bf{H}}^H{\bf W}^H{\bf{y}}_m\right)^H\right) \!\!.
\
\end{aligned}
\end{equation}

Since the ${\bf{y}}_m  $ is fixed received signal, the (\ref{eq4.1}) above is equal to the problem as
\begin{equation}\label{eq4.5}
\begin{aligned}
\mathop{\min} \limits_{r, \theta, \varphi} G(r, \theta, \varphi) 
= \sum_{m=1}^{M} {\bf{p}}_m^H{\bf{H}}^H{\bf W}^H{\bf W}{\bf{H}}{\bf{p}}_m\\
-\sum_{m=1}^{M}\left( {\bf{p}}_m^H{\bf{H}}^H{\bf W}^H{\bf{y}}_m+\left( {\bf{p}}_m^H{\bf{H}}^H{\bf W}^H{\bf{y}}_m\right)^H\right)
\end{aligned}
\end{equation}

The objective function $  G(r, \theta, \varphi)  $ can be optimized with an  iterative gradient descent approach methods. In the $ i $-th iteration, we need to calculate new estimated the three parameters, i.e., $ \hat{r}^{(i+1)} $, $ \hat{\theta}^{(i+1)} $, and $ \hat{\varphi}^{(i+1)} $, which can be updated as
\begin{equation}\label{eq4.9}
\begin{aligned}
\hat{{r}}^{(i+1)}&=\hat{{r}}^{(i)}-\eta_{r} \cdot \nabla_{{r}} G_{\mathrm{opt}}^{(i)}\left(\hat{{r}}^{(i)}, \hat{{\theta}}^{(i)}, \hat{{\varphi}}^{(i)}\right),
\end{aligned}
\end{equation}
\begin{equation}\label{eq4.9.1}
\begin{aligned}
\hat{{\theta}}^{(i+1)}&=\hat{{\theta}}^{(i)}-\eta_{\theta} \cdot \nabla_{{\theta}} G_{\mathrm{opt}}^{(i)}\left(\hat{{r}}^{(i)}, \hat{{\theta}}^{(i)}, \hat{{\varphi}}^{(i)}\right),
\end{aligned}
\end{equation}
\begin{equation}\label{eq4.9.2}
\begin{aligned}
\hat{{\varphi}}^{(i+1)}&=\hat{{\varphi}}^{(i)}-\eta_{\varphi} \cdot \nabla_{{\varphi}} G_{\mathrm{opt}}^{(i)}\left(\hat{{r}}^{(i)}, \hat{{\theta}}^{(i)}, \hat{{\varphi}}^{(i)}\right),
\end{aligned}
\end{equation}
where the gradients can be calculated according to Appendix A. $ \eta_{r} $, $ \eta_{\theta} $, $ \eta_{\varphi} $ denote the lengths of step for the distance and angles to guarantee the $ G_{\mathrm{opt}}^{(i)}\left(\hat{{r}}^{(i+1)}, \hat{{\theta}}^{(i+1)}, \hat{{\varphi}}^{(i+1)}\right) \leq G_{\mathrm{opt}}^{(i)}\left(\hat{{r}}^{(i)}, \hat{{\theta}}^{(i)}, \hat{{\varphi}}^{(i)}\right)  $. As the number of iterations increases, the parameters estimates becomes more accurate. 
{\color{black}Since only one variable in each iteration is optimized while the other variables remain unchanged, the gradient descent method utilized in this paper can be regarded as block coordinate descent.}
{\color{black}The iteration process will suspend until the biggest iteration number achieves or the normalized difference between the latest estimated parameters $ \left(\hat{{r}}^{(i+1)}, \hat{{\theta}}^{(i+1)}, \hat{{\varphi}}^{(i+1)}\right) $ and the last estimated parameters $ \left(\hat{{r}}^{(i)}, \hat{{\theta}}^{(i)}, \hat{{\varphi}}^{(i)}\right) $ are smaller than $ \epsilon $.}

Based on the $ \left(\hat{{r}}^{(I)}, \hat{{\theta}}^{(I)}, \hat{{\varphi}}^{(I)}\right) $, we can obtain the estimated  ${\hat{{\bf{H}}}}_{{\rm{LoS}}}$ by (\ref{eq3.5}). Then, we can eliminate the influence of ${\hat{{\bf{H}}}}_{{\rm{LoS}}}$ on the received pilots $ \bf{Y} $ and then estimate the ${\hat{{\bf{H}}}}_{{\rm{NLoS}}}$, which is shown as follows.  


\begin{algorithm}[htbp]
	\caption{NLoS path components estimation}
	\textbf{Inputs}: ${\bf{Y}}_{{\rm{NLoS}}}$, ${\bf{P}}$, ${\bf{W}}$, ${\bf{D}}_r$, ${\bf{D}}_t$, $L$.
	\\\textbf{Initialization}: ${\Omega_1}=\emptyset$,  ${\Omega_2}=\emptyset$, ${\Omega}=\emptyset$, ${\bf A}= {\bf{0}}_{S_1S_2\times 1}$, ${\bf{R}}={\bf{Y}}_{{\rm{NLoS}}}$, $ {\bf A}_t = {\bf D}_t^H{\bf P}$, $ {\bf A}_r = {\bf W}{\bf D}_r$, 
	\\1. \textbf{for} $l = 1,2,\cdots, L$ \textbf{do}
	\\2. \hspace*{+3mm}${n^{*}}={\mathop{\rm{argmax}}}\|{\rm vec}({{\bf A}_r}^H{{\bf{R}}{{\bf A}_t}^H)\|}^2_2$
	\\3. \hspace*{+3mm}$ n_1 = {\rm floor}(({n}^{*}-1)/N_2)+1$, $ n_2=\mod({n}^{*}-1,N_2)+1$
	\\4. \hspace*{+3mm}${\Omega}={\Omega}\bigcup {n}^{*}$, ${\Omega_1}={\Omega}\bigcup n_1$, ${\Omega_2}={\Omega}\bigcup n_2$
	\\5. \hspace*{+3mm}$ {\bf A }= \left[ {\bf A} \ {\bf A}_t\left(\Omega_1,:\right)^H\otimes   {\bf A}_r\left(:,\Omega_2\right)\right] $	
	\\6. \hspace*{+3mm}${\hat{{\bf{h}}}}_A({\Omega})=({\bf A}^H{\bf A})^{-1}{\bf A}^H{\rm vec}({\bf Y}_{\rm NLoS})$
	\\7. \hspace*{+3mm}reshape $ {\hat{{\bf{h}}}}_A $ into $ {\hat{{\bf{H}}}}_A $ of size $ S_2 \times S_1 $
	\\8. \hspace*{+3mm}${\bf{R}}={\bf{Y}}_{{\rm{NLoS}}}-{\bf{A}}_r{\hat{{\bf{H}}}}_A{\bf{A}}_t$
	\\9. \textbf{end for}
	\\10. ${\hat{{\bf{H}}}}_{{\rm{NLoS}}} = {\bf{D}}_r{\hat{{\bf{H}}}}_A{\bf{D}}_t^H $
	\\\textbf{Output}: Estimated NLoS path components ${\hat{{\bf{H}}}}_{{\rm{NLoS}}}$.
\end{algorithm}\vspace*{0mm}

\vspace*{0mm}
\subsection{Stage 2: NLoS Path Components Estimation}
As we already acquire ${\hat{{\bf{H}}}}_{{\rm{LoS}}}$, the received pilots without the effect of ${\hat{{\bf{H}}}}_{{\rm{LoS}}}$ can be presented as 
\begin{equation}\label{eq4.6}
{\bf{Y}}_{{\rm{NLoS}}} = {\bf{Y}}-{\bf W}{\hat{{\bf{H}}}}_{{\rm{LoS}}}{\bf{P}}.
\end{equation}
On the base of the polar-domain representation (\ref{eq9.1}), the $ {\bf{Y}}_{{\rm{NLoS}}} $ can be presented as
\begin{equation}\label{eq4.7}
{\bf{Y}}_{{\rm{NLoS}}} = {\bf W}{{\bf{H}}}_{{\rm{NLoS}}}{\bf{P}}+{\bf{N}} = {\bf W}{\bf{D}}_r{{\bf{H}}}_{{\rm{NLoS}}}^P{\bf{D}}_t^H{\bf{P}}+{\bf{N}}.
\end{equation}

As mentioned above, $  {\bf{H}}_{{\rm{NLoS}}}^P $ is sparse in polar-domain, thus the NLoS path components estimation is reformulated as a sparse recovery problem.  
In this sparse recovery problem, the NLoS path components sensing matrix at transmitter and receiver sides can be denoted as ${\bf{A}}_t  ={\bf{D}}_t^H{\bf{P}}$ and ${\bf{A}}_r ={\bf W}{\bf{D}}_r$. {\color{black}The low-complexity matrix based OMP algorithm to solve this problem can be summarized in \textbf{Algorithm 2}. }

Specifically, since there are $ L $ components in the polar-domain, we will conduct $ L $ iterations to find $ L $ supports in transmitter sensing matrix ${\bf{A}}_t $ and  $ L $ supports in receiver sensing matrix $ {\bf{A}}_r $. In $ l $-th iteration, we will calculate the correlation between the transmitter and receiver sensing matrices ${\bf{A}}_t $, ${\bf{A}}_r $ and the residual matrix $ \bf{R} $. 
In Step 4, we obtain the updated support $ \Omega $, $ \Omega_1 $, $ \Omega_2 $, where $ \Omega_1 $, $ \Omega_2 $ denote the support of transmitter and receiver sides. Then, in Step 6, the currently estimated near-field NLoS path component $ {\hat{{\bf{h}}}}_A $ is calculated by the least square (LS) algorithm. In Step 7, the $ {\hat{{\bf{h}}}}_A $ need to be reshaped into $ {\hat{{\bf{H}}}}_{A} $ in
the polar-domain of the size $ S_2 \times S_1 $. Finally, after $ L $ iterations are performed, we obtain the estimated NLoS path components ${\hat{{\bf{H}}}}_{{\rm{NLoS}}}$. {\color{black}Since there exists an estimation error in the first stage, i.e., LoS path component estimation, the error of NLoS path components estimation will contain the error of LoS path component estimation.}


After estimating the near-field LoS path component and the NLoS paths components, the $ {\hat{{\bf{H}}}}  $ can be written as 
\begin{equation}
{\hat{{\bf{H}}}}  ={\hat{{\bf{H}}}}_{{\rm{LoS}}}+{\hat{{\bf{H}}}}_{{\rm{NLoS}}}.
\end{equation}

\vspace*{0mm}
\subsection{Cram\'er-Rao Lower Bound}
The CRLB bound is a theoretical bound of MSE to evaluate channel estimation algorithms~\cite{CRLB}. For the channel estimation problem~(\ref{eq1}), we utilize $ {\rm{vec}}({\bf{ABC}})=({\bf{C}}^T \otimes {\bf{A}}){\rm{vec}}({\bf{B}}) $ to reformulate (\ref{eq1}) as

\begin{equation}\label{eq4.8}
{\bf{y}} = ({\bf{P}}^T \otimes {\bf{W}}){\bf{h}}+{\bf{n}} =  {\bf{ Q}} {\bf{h}}+{\bf{n}}
\end{equation}
where $ {\bf{y}} ={\rm{vec}}({\bf{Y}}) \in \mathbb{C}^{{N_{\rm r}^{\rm RF}}M\times 1}$, ${\bf{h}} = {\rm{vec}}({{\bf{H}}})\in \mathbb{C}^{N_1N_2\times 1}$, ${\bf{n}}= {\rm{vec}}({\bf{N}}) \in \mathbb{C}^{{N_{\rm r}^{\rm RF}}M\times 1} $, and ${\bf{Q}}= ({\bf{P}}^T \otimes {\bf{W}}) \in \mathbb{R}^{{N_{\rm r}^{\rm RF}}M\times N_1N_2} $. Since $ \bf Q $ is a real matrix and $ \bf y,h,n $ are complex vectors, we can split problem (\ref{eq4.8}) into real part and imaginary part as
\begin{equation}\label{eq4.8.1}
\begin{aligned}
{\bf{y}}_u =  {\bf{ Q}} {\bf{h}}_u+{\bf{n}}_u\\
{\bf{y}}_v =  {\bf{ Q}} {\bf{h}}_v+{\bf{n}}_u
\end{aligned}
\end{equation}
where $ {\bf{y}}_u = {\rm Re}({\bf{y}}) $, $ {\bf{y}}_v = {\rm Im}({\bf{y}}) $, $ {\bf{h}}_u = {\rm Re}({\bf{h}}) $, $ {\bf{h}}_v = {\rm Im}({\bf{h}}) $, $ {\bf{n}}_u = {\rm Re}({\bf{n}}) $, $ {\bf{n}}_v = {\rm Im}({\bf{n}}) $. We denote $\hat{ {\bf h} } = \hat{ {\bf h} }_u+\hat{ {\bf h} }_v$ as the estimated channel.
{\color{black}The CRLB of the unbiased estimator $ {\hat{\bf{h}}}$ can also be divided into two parts, which are shown as}
\begin{equation}
\begin{aligned}
{\rm CRLB}&={\rm CRLB}_u+{\rm CRLB}_v\\
&=\mathrm{E}\left\{\left\|\widehat{\mathbf{h}}-\mathbf{h}\right\|^2\right\} \\
&= \mathrm{E}\left\{\left\|\widehat{\mathbf{h}}_u-\mathbf{h}_u\right\|^2\right\} +\mathrm{E}\left\{\left\|\widehat{\mathbf{h}}_v-\mathbf{h}_v\right\|^2\right\}
\end{aligned}
\end{equation}

In this case, we first consider real part of (\ref{eq4.8.1}). Since the ${ \bf n}_u $ follows the distribution of Gaussian distribution with $ \bf{0} $ mean and $ {\bf{\sigma}}^2 $ variance, the conditional probability density function of $ {\bf y}_u $ with the given $ {\bf h}_u $ is
\begin{equation}
\begin{aligned}
p_{\mathbf{y}_u \mid \mathbf{h}_u} \left(\mathbf{y}_u ; \mathbf{h}_u
\right) 
=\frac{1}{\left(2 \pi \sigma^2\right)^{{N_{\rm r}^{\rm RF}}M / 2}} \exp \left\{-\frac{1}{2 \sigma^2}\left\|{\bf{y}}_u - {\bf{ Q}} {\bf{h}}_u\right\|^2\right\}
\end{aligned}.
\end{equation}
The Fisher information matrix of real part of (\ref{eq4.8.1}) can then be derived as
\begin{equation}
\begin{aligned}
{[\mathbf{J}]_{m, n} } & \triangleq-\mathrm{E}\left\{\frac{p_{\mathbf{y}_u \mid \mathbf{h}_u} \left(\mathbf{y}_u ; \mathbf{h}_u\right)}{\partial h_{u, m} \partial h_{u, n}}\right\} 
&=\frac{1}{\sigma^2}\left[\mathbf{Q}^H \mathbf{Q}\right]_{m, n},
\end{aligned}
 \end{equation}
 where $  h_{u, m} , h_{u, n} $ denote the $ m $-th and $ n $-th entry of $ \mathbf{h}_u $. Then, the real part $ {\rm CRLB}_u  $ is
\begin{equation}\label{eq4.8.2}
\begin{aligned}
{\rm CRLB}_u &=\mathrm{E}\left\{\left\|\widehat{\mathbf{h}}_u-
\mathbf{h}_u\right\|^2\right\} \geq  {\rm Tr}\left\lbrace  {\bf J}_u^{-1}\right\rbrace = \sigma^2 {\rm Tr}\left\lbrace({\bf Q}^H{\bf Q})^{-1} \right\rbrace .
\end{aligned}
\end{equation}
Since ${\bf{Q}}= {\bf{P}}^T \otimes {\bf{W}} $, $ ({\bf Q}^H{\bf Q})^{-1}  $ can be presented as 
\begin{equation}\label{eq4.8.3} 
\begin{aligned}
({\bf Q}^H{\bf Q})^{-1} &= \left(  ({\bf{P}}^T \otimes {\bf{W}})^H ({\bf{P}}^T \otimes {\bf{W}})\right) ^{-1}\\
&= \left( \left({\bf P}{\bf P}^H \right)^T \otimes \left({\bf W}^H{\bf W} \right)  \right) ^{-1}\\
&= \left( \left({\bf P}{\bf P}^H \right)^{-1}\right) ^T \otimes \left({\bf W}^H{\bf W} \right) ^{-1}.
\end{aligned}
\end{equation}
Thus, $ {\rm Tr}\left\lbrace({\bf Q}^H{\bf Q})^{-1} \right\rbrace $ can be calculated as
\begin{equation}\label{eq4.8.4} 
\begin{aligned}
{\rm Tr}\left( ({\bf Q}^H{\bf Q})^{-1}\right)  
&= {\rm Tr}\left(\left( \left({\bf P}{\bf P}^H \right)^{-1}\right) ^T\right)  {\rm Tr}\left( \left({\bf W}^H{\bf W} \right) ^{-1}\right)\\
&= {\rm Tr}\left(\left({\bf P}{\bf P}^H \right)^{-1}\right)  {\rm Tr}\left( \left({\bf W}^H{\bf W} \right) ^{-1}\right)\\
&= \left( \sum_{i=1}^{N_1} \lambda_i^{-1}\right) \left(  \sum_{j=1}^{N_2} \eta_j^{-1} \right) \\
&\geq {N_1}\left({N_1} / \sum_{i=1}^{N_1} \lambda_i\right) {N_2}\left({N_2} / \sum_{j=1}^{N_2} \eta_j\right)  \\
&=\frac{{N_1}^2}{\operatorname{Tr}\left\{{\bf P}{\bf P}^H\right\}} \frac{{N_2}^2}{\operatorname{Tr}\left\{{\bf W}^H\bf W\right\}}.
\end{aligned}
\end{equation}

$ {\left\lbrace \lambda_i\right\rbrace}_{i=1}^{N_1}  $ and $ {\left\lbrace \eta_j\right\rbrace}_{j=1}^{N_2}  $ are denoted as the $ N_1 $  and $ N_2 $ eigenvalues of the matrix of $ {\bf P}{\bf P }^H$ and $ {\bf W}^H\bf W $. 
The equality in (\ref{eq4.8.4}) holds when these eigenvalues satisfy $ \lambda_1 =\lambda_2=\cdots\lambda_{N_1} $, and $ \eta_1 =\eta_2=\cdots\eta_{N_2} $, i.e.,  the columns of $ {\bf P}^H $ ($ \bf W $) and are orthogonal. Thus, the matrix $ {\bf P}{\bf P }^H $ of size $ N_1 \times N_1 $ has identical diagonals equal to $ M $, and $ {\bf W}^H{\bf W } $ of size $ N_2 \times N_2 $ has identical diagonals equal to 
$ N^{\rm RF}_{\rm r} $. In this case, $ {\rm Tr}\left\lbrace({\bf P}{\bf P})^H \right\rbrace {\rm Tr}\left\lbrace({\bf W}^H\bf W) \right\rbrace = N_1MN_2N^{\rm RF}_{\rm r}$. Finally, the CRLB of the real part of our estimation problem becomes
\begin{equation}\label{eq4.8.5}
\begin{aligned}
{\rm CRLB}_u &=\mathrm{E}\left\{\left\|\widehat{\mathbf{h}}_u-
\mathbf{h}_u\right\|^2\right\} = \sigma^2 \dfrac{N_1N_2}{MN^{\rm RF}_{\rm r}}  .
\end{aligned}
\end{equation}

From the (\ref{eq4.8.1}), we can observe that the real part and imaginary part have the same form, $ {\rm CRLB}_v = {\rm CRLB}_u = \sigma^2 \dfrac{N_1N_2}{MN^{\rm RF}_{\rm r}} $. At last, thus the CRLB of the (\ref{eq4.8}) is
\begin{equation}\label{CRLB}
 {\rm CRLB} =  {\rm CRLB}_u + {\rm CRLB}_v = 2\sigma^2 \dfrac{N_1N_2}{MN^{\rm RF}_{\rm r}}.
\end{equation}

It is noteworthy that the columns of pilot matrix $ {\bf P}^H $ and combing matrix $ {\bf W} $ are not orthogonal in practical. Thus, the MSE of the practical channel estimation algorithm cannot achieve the CRLB bound in (\ref{CRLB}), which is verified in the simulation results in Section~\ref{S6}.

\vspace*{3mm}
\subsection{Computational Complexity Analysis}
For the proposed two stage channel estimation algorithm, we analyze its computational complexity as follows. In the stage of the LoS path component estimation, we can observe that the complexity comes from two parts, i.e., coarse on-grid estimation and refining processes. {\color{black} In Steps 1-4, for coarse on-grid estimation, we need to compute the $ {\bf{H}}^*_{{\rm{LoS}}} $ according to (\ref{eq3.5}) in parameters collection and find the best parameters. Therefore, the complexity of this part is $ \mathcal{O}\left(S_{{\rm{LoS}}}\left( N_{\rm r}^{\rm RF}N_1N_2+N_{\rm r}^{\rm RF}N_1M\right)\right)  $, where $ S_{{\rm{LoS}}} $ is the size of the parameters collection. Since we only need a coarse on-grid estimation, $ S_{{\rm{LoS}}} $ usually is small. Then, for the refining process in Steps 5-9, the complexity is introduced by the gradient calculation. The complexity to calculate gradients $ \mathcal{O}(MI(N_{\rm r}^{\rm RF}N_1N_2+N_{\rm r}^{\rm RF}N_1+N_{\rm r}^{\rm RF}N_2+N_2N_1+N_{\rm r}^{\rm RF})) $. 
Since the $ N_1$, 
$N_2$ is usually much larger than $N_{\rm r}^{\rm RF}  $ and $M$, the complexity of the gradient calculation can be presented as $ \mathcal{O}\left(\left( S_{{\rm{LoS}}}+MI\right) N_{\rm r}^{\rm RF}N_1N_2\right)   $.}
For the NLoS path components estimation, the computational complexity can be obtained as $ \mathcal{O}(N_1N_2(S_1+S_2)L) $ by referring to the OMP algorithm~\cite{OMP}. 

\vspace{+0mm}
\section{Simulation Result}\label{S6}
In this section, we conduct the simulations to verify the performance of the proposed two stage channel estimation algorithm for the proposed mixed LoS/NLoS near-field XL-MIMO channel model. 
The system parameters are as follows: the number of antenna of transmitter is $ N_1 = 256 $, the number of antenna of receiver is $ N_2 = 128 $. The carrier frequency is $ f = 50 $ GHz, corresponding to $ \lambda = 0.006 \,{\rm m} $. By utilizing (\ref{MIMO_RD}), the MIMO-RD can be calculated as $ \frac{2(D_1+D_2)^2}{\lambda} = \frac{2({N_1\lambda/2}+N_2\lambda/2)^2}{\lambda} = 442.7\,{\rm m} $ in this scenario. By utilizing (\ref{MIMO_ERD}), the MIMO-ARD can be calculated as $ \frac{4D_1D_2}{\lambda} = \frac{4({N_1\lambda/2})(N_2\lambda/2)}{\lambda} = 196.6\,{\rm m} $ in this scenario. The near-field channel in (\ref{eq3.1}) contains $  L = 3 $ NLoS path components. Meanwhile, the sampled angles of arrival follow the uniform distribution $ \mathcal{U}\left(-\frac{\pi}{3}, \frac{\pi}{3}\right) $. The distances are generated in range of [50, 500] meters. The pilot matrix $ \bf{P} $ and combining matrix $ \bf W $ randomly chooses their elements from $ \left\{-\frac{1}{\sqrt{M}},+\frac{1}{\sqrt{M}}\right\} $ and $ \left\{-\frac{1}{\sqrt{N_2}},+\frac{1}{\sqrt{N_2}}\right\} $~\cite{pilot}. 


\begin{table}[H]\footnotesize
	\begin{center}
		\vspace{-3mm}
		\caption{\color{black}Complexity Comparison }	
		\vspace{-3mm}
		\begin{tabular}{|c|c|}
			\hline
			\color{black}Method                                          & \color{black}Complexity \\ \hline
			\color{black}\begin{tabular}[c]{@{}c@{}}Far-field codebook   \\ based OMP\end{tabular}      & \color{black}$ \mathcal{O}(N_1N_2(N_1+N_2)L) $          \\ \hline
			\color{black}\begin{tabular}[c]{@{}c@{}}Near-field codebook   \\ based OMP\end{tabular}      & \color{black}$ \mathcal{O}(N_1N_2(S_1+S_2)L) $         \\ \hline
			\color{black}\begin{tabular}[c]{@{}c@{}}Proposed two stage  \\ channel estimation\end{tabular}   & \color{black}$ \mathcal{O}\!\left(\left( S_{{\rm{LoS}}}\!+\!MI\right)  N_{\rm r}^{\rm RF}N_1N_2\!+\!N_1N_2(S_1\!+\!S_2)L\right) $    \\ \hline
		\end{tabular}\label{Table1}
	\end{center}
	\vspace{-3mm}
\end{table}

We compare the proposed two stage channel estimation algorithm and the existing near-field codebook based OMP method~\cite{Cui_Tcom} and far-field codebook based OMP method~\cite{OMP}. 
{\color{black}The complexities of the three methods are presented in Table~\ref{Table1}.} The normalized mean square error (NMSE) performance, which is defined as $ \mathrm{NMSE}=\frac{\mathbb{E}(\|\mathbf{H}-\hat{\mathbf{H}}\|_{2}^{2})}{\mathbb{E}(\|\mathbf{H}\|_{2}^{2})}$, is used to evaluate the accuracy of different methods. It worth to point out that we use the $ \frac{\rm CRLB}{{\mathbb{E}(\|\mathbf{H}\|_{2}^{2})}} $ as the bound of NMSE performance.

\begin{figure}[tbhp]
	\begin{center}
		\vspace*{0mm}\includegraphics[width=1\linewidth]{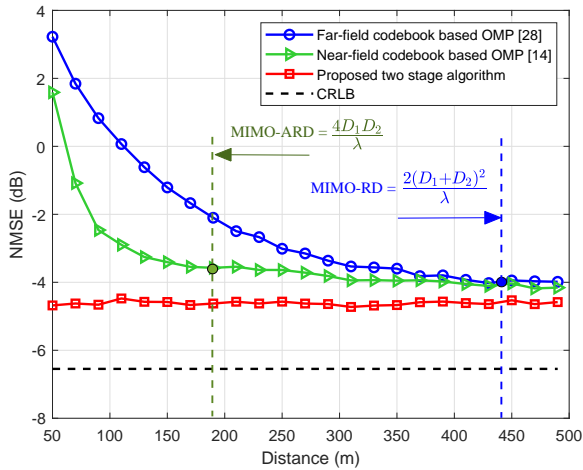}
	\end{center}
	\vspace*{-4mm}\caption{NMSE performance comparison with respect to the distance of the transmitter from the receiver.}\vspace{+0mm} \label{FIG2}
\end{figure}

Fig.~\ref{FIG2} depicts the NMSE performance comparison with respect to the distance of the transmitter from the receiver. The range of distance is from 50 ${\rm m} $ to 500 ${\rm m} $. The SNR is 5dB and the size of pilot matrix is $ 256 \times 128 $, the compressive ratio is $ \dfrac{M}{N_1}=0.5 $~\cite{comratio}. The proposed two stage method can achieve better NMSE performance than the existing far-field codebook based OMP method and the near-field codebook based OMP method. Specifically, the NMSE of the proposed algorithm is robust and remains the lowest value of all the schemes in the whole range of distance. The NMSE performance of the far-field codebook based OMP method and the near-field codebook based OMP method degrades gradually with the decrease of the distance. The reason is that there are still phase discrepancies between the proposed channel and the existing far-field array response vectors based channel as well as the existing near-field channel. In other words, the existing two channel models mismatch the feature of the piratical near-field XL-MIMO scenario. 
Furthermore, we can observe that the MIMO-RD can capture the turning point of performance loss between the proposed two stage scheme and the far-field codebook based OMP scheme. Specifically, when the distance is larger than $442.7\,{\rm m}$, the performance of the proposed two stage scheme and the far-field codebook based OMP channel estimation show the same performance. The reason is that when the distance is larger than MIMO-RD, the proposed channel model degenerates into the far-field channel model. Similarly, MIMO-ARD can capture the turning point of performance loss between the proposed two stage scheme and the near-field codebook based OMP scheme. When the distance is larger than MIMO-ARD, i.e., $196.6\,{\rm m}$, the performance of the proposed two stage scheme and the near-field codebook based OMP channel estimation show the same performance. This is because when the distance is larger than MIMO-ARD, the phase discrepancy between the proposed channel model and the existing model vanishes. 

\begin{figure}[tp]
	
	\vspace*{0mm}
	\centering 
	\subfigure[]{
		
		\vspace*{0mm}\includegraphics[width=1\linewidth]{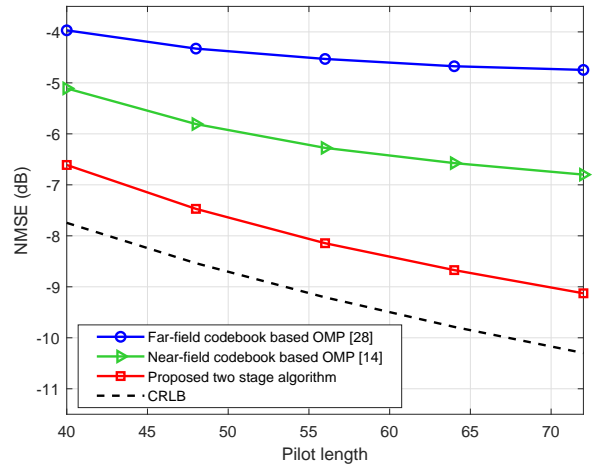}
		
	}\vspace*{-2mm}
	
	\subfigure[]{
		
		\vspace*{0mm}\includegraphics[width=1\linewidth]{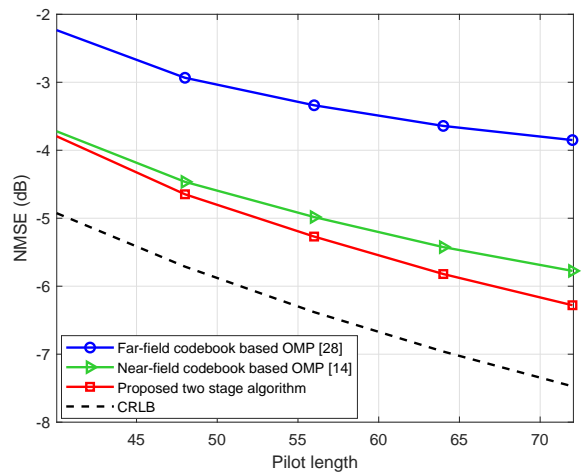}
	}
	\vspace*{0mm}
	\caption{ \color{black} NMSE performance with respect to the size of the pilot. (a) $ r = 60\,{\rm m}$; (b) $ r = 100\,{\rm m}$.}\vspace{0mm}
	\label{FIG4}
\end{figure}

\vspace{2mm}
Fig.~\ref{FIG4} depicts the NMSE performance comparison with respect to the size of the pilot $ \bf{P} $ under the condition of 5 dB SNR. 
The proposed scheme achieves the best performance out of all three considered schemes. When the distance becomes large as shown in~Fig.~\ref{FIG4} (b), the NMSE achieved by the two stage channel estimation scheme and other schemes are similar. {\color{black}Since the fewer pilots, the greater the error in the first stage of the LoS path component, the gap between near-field codebook-based OMP and the proposed algorithm at low pilot overhead is smaller than that of low pilot overhead. When the distance becomes smaller as shown in Fig.~\ref{FIG4} (a), the proposed scheme outperforms all the other schemes.}

\begin{figure}[tp]
	
	\vspace*{0mm}
	\centering 
	\subfigure[]{
		
		\vspace*{0mm}\includegraphics[width=1\linewidth]{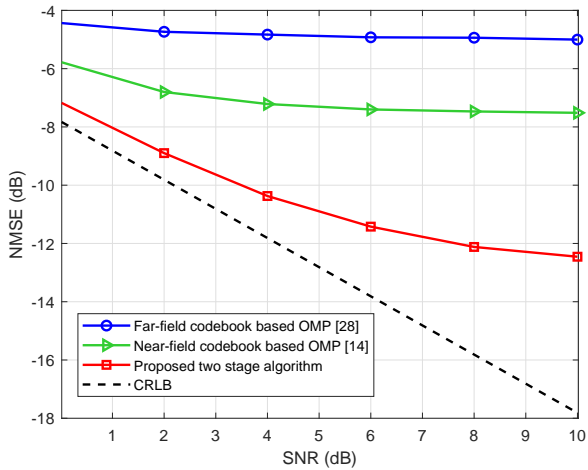}
		
	}\vspace*{-4mm}
	
	\subfigure[]{
		
		\vspace*{0mm}\includegraphics[width=1\linewidth]{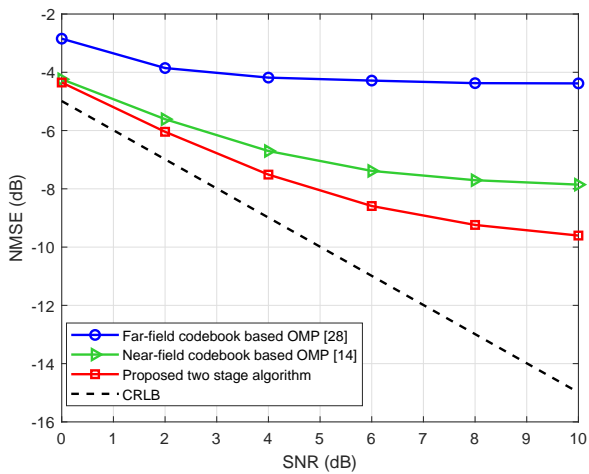}
	}
	\vspace*{0mm}
	\caption{ \color{black}NMSE performance comparison with respect to the SNR under different distances. (a) $ r = 60\,{\rm m} $; (b) $ r = 100\,{\rm m} $.}\vspace{+0mm}
	\label{FIG3}
\end{figure}

Fig.~\ref{FIG3} shows the NMSE performance comparison with respect to the SNR under different distances, where the size of pilot is $ 256 \times 64 $. In Fig.~\ref{FIG3} (a) and (b), the distance is $60\,{\rm m}$ and $100\,{\rm m}$, respectively. We can see that when the distance is smaller than MIMO-ARD, the proposed two stage channel estimation scheme outperforms all the other schemes. In particular, when SNR is 5 dB and $r=60\,{\rm m}$, the proposed scheme can achieve about 4 dB improvement compared with the near-field codebook based OMP scheme. The reason is that the existing far-field and near-field codebook based OMP schemes cannot deal with the LoS path component of near-field XL-MIMO channel model when the distance is smaller than MIMO-ARD. Furthermore, we can see that, in the Fig.~\ref{FIG3} (b), the performance gap of NMSE of the three schemes become smaller in the range of all the SNRs. 

{\color{black}In order to evaluate the proposed two stage algorithm in the realistic datasets, we utilize QuaDRiGa channel emulation platform~\cite{QuaDRiGa1,QuaDRiGa2} as the channel generator to test our proposed method. The standard channel cluster model of 3GPP TR 38.901~\cite{3GPP} is used to generate the channel dataset. The main parameters of the channel are set as shown in the Table.~\ref{par}.}

\vspace{0mm}	
\begin{table}[H]
	\begin{center}
		\vspace{0mm}
		\begin{center}
			\caption{\color{black}Parameters of 3GPP TR 38.901 channel }\label{par}	
		\end{center}	
		\vspace{-3mm}
		\footnotesize
		\begin{tabular}{|c|c|}
			\hline
			{\color{black}Parameters}                     & {\color{black}Value}                                                                     \\ \hline
			{\color{black}Simulation scenario }           & {\color{black}UMa}                                                                       \\ \hline
			{\color{black}Frequency}                      & {\color{black}3.5GHz}                                                                    \\ \hline
			{\color{black}Number of antennas at transmitter} & {\color{black}64}                                                                       \\ \hline
			{\color{black}Number of antennas at receiver}    & {\color{black}256}                                                                        \\ \hline
			{\color{black}Receiver location}              & \begin{tabular}[c]{@{}c@{}}{\color{black}Outdoor and Indoor}\\ {\color{black}LoS and NLoS}\end{tabular} \\ \hline
		\end{tabular}
	\end{center}
	\vspace{-3mm}
\end{table}

\begin{figure}[hbtp]
	\begin{center}
		\vspace*{0mm}\includegraphics[width=1\linewidth]{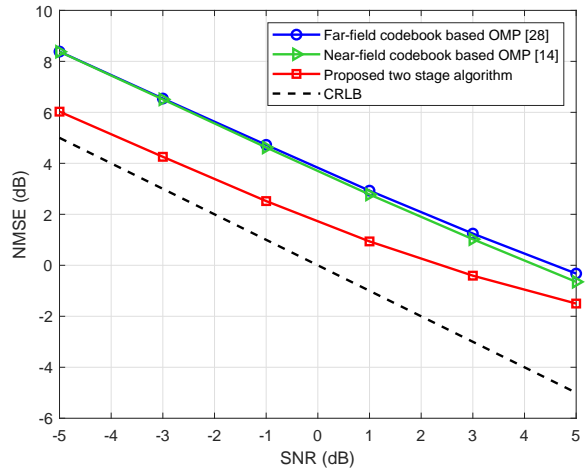}
	\end{center}
	\caption{\color{black}NMSE performance comparison with respect to the SNR under the QuaDRiGa channel dataset.}\label{GUA}
\end{figure}
\vspace*{0mm}

{\color{black}The specific simulation result with the GuaDRiGa channel dataset is shown in Fig.~\ref{GUA}. It can be observed that the proposed method still outperforms the existing far-field codebook-based OMP algorithm and the near-field codebook based OMP algorithm. Thus, the proposed mixed LoS/NLoS near-field MIMO channel model is more accurate compared with the existing model. Furthermore, the proposed two stage channel estimation scheme is an efficient scheme for the practical near-field XL-MIMO.} 

 
\vspace{0mm}
\section{Conclusions}\label{S7}
In this paper, the channel estimation of the near-field XL-MIMO scenario was investigated. We proposed the mixed LoS/NLoS near-field XL-MIMO channel model, where the LoS and NLoS path components were characterized by geometric free space assumption and the near-field response vectors, respectively. Then, we derived the range of the near-field region of XL-MIMO, i.e., MIMO-RD and MIMO-ARD. 
It is worth pointing out that the mixed LoS/NLoS near-field XL-MIMO channel model proposed in this paper can be recognized as the generalization of both the far-field channel model as well as the existing near-field channel model for MIMO systems. 
{\color{black}Simulation results showed that, compared with the far-field codebook based and near-field codebook based channel estimation schemes, the proposed two stage channel estimation scheme achieved better NMSE performance in both theoretical and practical channel models.}  For future work, the uniform planar array (UPA) based 3D near-field XL-MIMO channel model will attract further research interest. 



\vspace{0mm}
\begin{appendices}
    \section{Optimization of $G^{(i)}$ in with regard to $r^{(i)}$, $\theta^{(i)}$, $\varphi^{(i)}$}\label{AppendixA}	
In this appendix, the derivations of the $G^{(i)}$ in with regard to $r^{(i)}$, $\theta^{(i)}$, $\varphi^{(i)}$ are provided. For the simplification of expression, we will ignore the superscript $ (i) $.
  
For the distance $ r $, the gradient of $G$ is given by
  \begin{equation}\label{app1}
    \begin{aligned}
    \frac{{\partial {G}}}{{\partial {r}}} =&  \sum_{m=1}^{M}{\bf{p}}_m^H\frac{\partial {\bf{H}}^H}{{\partial {r}}}{\bf W}^H{\bf W}{\bf{H}}{\bf{p}}_m+
    \sum_{m=1}^{M}{\bf{p}}_m^H{\bf{H}}^H{\bf W}^H{\bf W}\frac{\partial {\bf{H}}}{{\partial {r}}}{\bf{p}}_m\\
    &-\sum_{m=1}^{M} {\bf{p}}_m^H\frac{\partial {\bf{H}}^H}{{\partial {r}}}{\bf W}^H{\bf{y}}_m-\sum_{m=1}^{M} {\bf{y}}_m^H{\bf W}\frac{\partial {\bf{H}}}{{\partial {r}}}{\bf{p}}_m.
    \end{aligned}
    \vspace{-1.5mm}
    \end{equation}
Since $ r $ is a real variable, $ \frac{\partial {\bf{H}}^H}{{\partial {r}}} = \left( \frac{\partial {\bf{H}}}{{\partial {r}}} \right)^H$. Thus, the (\ref{app1}) can be rewritten as
\begin{equation}\label{app2}
\begin{aligned}
\frac{{\partial {G}}}{{\partial {r}}} =\sum_{m=1}^{M} {\bf{p}}_m^H\left( \frac{\partial {\bf{H}}}{{\partial {r}}} \right)^H{\bf W}^H\left( {\bf W}{\bf{H}}{\bf{p}}_m-{\bf{y}}_m \right)
\\+ \sum_{m=1}^{M}\left( {\bf{p}}_m^H{\bf{H}}^H{\bf W}^H-{\bf y}_m^H\right) {\bf W}\frac{\partial {\bf{H}}}{{\partial {r}}}{\bf{p}}_m.
\end{aligned}
\vspace{-1.5mm}
\end{equation}

Similar to the derivation of $ \frac{{\partial {G}}}{{\partial {r}}} $ above, the $ \frac{{\partial {G}}}{{\partial {\theta}}} $ and  $ \frac{{\partial {G}}}{{\partial {\varphi}}} $ can be presented as
\begin{equation}\label{app3}
\begin{aligned}
\frac{{\partial {G}}}{{\partial {\theta}}} 
= \sum_{m=1}^{M} {\bf{p}}_m^H\left( \frac{\partial {\bf{H}}}{{\partial {\theta}}} \right)^H{\bf W}^H\left( {\bf W}{\bf{H}}{\bf{p}}_m-{\bf{y}}_m \right)\\
+ \sum_{m=1}^{M}\left( {\bf{p}}_m^H{\bf{H}}^H{\bf W}^H-{\bf y}_m^H\right) {\bf W}\frac{\partial {\bf{H}}}{{\partial {\theta}}}{\bf{p}}_m,
\end{aligned}
\vspace{-1.5mm}
\end{equation}

\begin{equation}\label{app4}
\begin{aligned}
\frac{{\partial {G}}}{{\partial {\varphi}}} 
= \sum_{m=1}^{M} {\bf{p}}_m^H\left( \frac{\partial {\bf{H}}}{{\partial {\varphi}}} \right)^H{\bf W}^H\left( {\bf W}{\bf{H}}{\bf{p}}_m-{\bf{y}}_m \right)
\\+ \sum_{m=1}^{M}\left( {\bf{p}}_m^H{\bf{H}}^H{\bf W}^H-{\bf y}_m^H\right) {\bf W}\frac{\partial {\bf{H}}}{{\partial {\varphi}}}{\bf{p}}_m.
\end{aligned}
\vspace{-1.5mm}
\end{equation}

Based on the $\left(n_2,n_1\right)  $-th element in the $ \bf H $, i.e., $ {\bf H}_{n_2,n_1}$ $ = $ $ {{\frac{1}{{r_{{n_2},{n_1}}}}}{e^{ - j2\pi {r_{{n_2},{n_1}}}/\lambda }}} $, where $ {r_{n_2,n_1}} $ is presented in (\ref{Rn1n2}), we can obtain the $\left(  n_2,\!n_1\right)  $-th element in the $ \frac{\partial {\bf H}}{{\partial {r}}} $, $ \frac{\partial {\bf H}}{{\partial {\theta}}} $, and $ \frac{\partial {\bf H}}{{\partial {\varphi}}} $ as
\begin{equation}\label{app6}
\begin{aligned}
\frac{{\partial {\bf H}_{n_2,n_1}}}{{\partial {r}}} 
&= \left( d_1\sin\theta-d_2\sin(\theta+\varphi)-r\right) \left( 1\!+\!j{r_{{n_2},{n_1}}}\right)\Gamma, 
\end{aligned}
\vspace{-1.5mm}
\end{equation}

\begin{equation}\label{app7}
\begin{aligned}
\frac{{\partial {\bf H}_{n_2,n_1}}}{{\partial {\theta}}} 
&= r\left( d_1\cos\theta-d_2\cos(\theta+\varphi)\right) \left( 1\!+\!j{r_{{n_2},{n_1}}}\right)\Gamma ,
\end{aligned}
\vspace{-1.5mm}
\end{equation}

\begin{equation}\label{app8}
\begin{aligned}
\frac{{\partial {\bf H}_{n_2,n_1}}}{{\partial {\varphi}}} 
&= jd_2\left( d_1\sin\varphi+r\cos(\theta+\varphi) \right)  \left({r_{{n_2},{n_1}}}\!-\!j\right) \Gamma
.\end{aligned}
\vspace{-1.5mm}
\end{equation}
where $\Gamma =\frac{e^{-j{r_{{n_2},{n_1}}}}}{{r_{{n_2},{n_1}}^3}}   $.

\end{appendices}
\vspace{5mm}

\bibliography{IEEEabrv,Gao1Ref}

\footnotesize

\bibliographystyle{IEEEtran}

\vspace{-1cm}

\end{document}